\documentclass[10pt, final, journal, letterpaper, twoside, twocolumn]{IEEEtran}
\usepackage{amsmath,amssymb}
\usepackage{graphicx}
\usepackage{physics}
\usepackage{optidef}
\usepackage{graphicx,epstopdf,epsfig}
\usepackage{adjustbox}
\usepackage[dvipsnames]{xcolor}
\usepackage[caption=false,font=footnotesize]{subfig}
\usepackage{siunitx}
\usepackage[linesnumbered,ruled]{algorithm2e}
\usepackage{float}
\usepackage{url}
\usepackage[normalem]{ulem}
\usepackage[section]{placeins} 
\usepackage{centernot}
\usepackage{enumerate}
\usepackage{textcomp}
\usepackage{diagbox}
\usepackage{hhline}
\usepackage{stackengine}
\usepackage[noadjust]{cite}

\newcommand{\Sy}{\mathbb{S}}
\renewcommand{\Re}{\mathbb{R}}
\newcommand{\Rc}{\overline{\Re}}

\newcommand{\Na}{\mathbb{N}}
\newcommand{\Xr}{\mathbb{X}}

\newcommand{\mcR}{\mathcal{R}}
\newcommand{\mcN}{\mathcal{N}}
\newcommand{\mcL}{\mathcal{L}}
\newcommand{\mcD}{\mathcal{D}}
\newcommand{\mcM}{\mathcal{M}}

\def\A{\mathbf{A}}

\def\C{\mathbf{C}}
\def\D{\mathbf{D}}
\def\G{\mathbf{G}}
\def\H{\mathbf{H}}
\def\I{\mathbf{I}}
\def\K{\mathbf{K}}
\def\LA{\mathbf{\Lambda}}
\def\M{\mathbf{M}}
\def\P{\mathbf{P}}
\def\Q{\mathbf{Q}}
\def\R{\mathbf{R}}
\def\S{\mathbf{S}}
\def\T{\mathbf{T}}
\def\U{\mathbf{U}}
\def\V{\mathbf{V}}
\def\W{\mathbf{W}}
\def\ZE{\mathbf{0}}

\def\a{\boldsymbol{a}}
\def\b{\boldsymbol{b}}
\def\d{\boldsymbol{d}}
\def\i{\boldsymbol{i}}
\def\j{\boldsymbol{j}}
\def\q{\boldsymbol{q}}
\def\u{\boldsymbol{u}}
\def\v{\boldsymbol{v}}
\def\w{\boldsymbol{w}}
\def\x{\boldsymbol{x}}
\def\y{\boldsymbol{y}}
\def\z{\boldsymbol{z}}

\def\bnu{\boldsymbol{\nu}}

\def\lambdamax{\lambda_{\mathrm{max}}}

\DeclareMathOperator*{\argmin}{\arg\!\min}
\DeclareMathOperator*{\prox}{\mathrm{prox}}
\DeclareMathOperator*{\minimize}{minimize}
\DeclareMathOperator*{\dom}{dom}
\DeclareMathOperator*{\epi}{epi}

\newcommand{\normH}[1]{\left\|#1\right\|_{\H}}
\newcommand{\normI}[1]{\left\|#1\right\|_2}

\graphicspath{{./Images/}}

\newtheorem{theorem}{Theorem}[section]

\newtheorem{definition}[theorem]{Definition}
\newtheorem{corollary}[theorem]{Corollary}
\newtheorem{lemma}[theorem]{Lemma}
\newtheorem{example}[theorem]{Example}
\newtheorem{remark}[theorem]{Remark}

\newcommand\xqed[1]{%
  \leavevmode\unskip\penalty9999 \hbox{}\nobreak\hfill
  \quad\hbox{#1}}
\newcommand{\fd}{\xqed{$\blacklozenge$}} 

\begin{document}

\title{On Plug-and-Play Regularization using \\ Linear Denoisers}

\author{Ruturaj~G.~Gavaskar, Chirayu~D.~Athalye, and Kunal~N.~Chaudhury, \IEEEmembership{Senior~Member, IEEE}

\thanks{The work of C.~D.~Athalye was supported by the Department of Science and Technology, Government of India under the Grant IFA17-ENG227. The work of K.~N.~Chaudhury was supported by MATRICS grant MTR/2018/000221 from the Department of Science and Technology, Government of India, and grant ISTC/EEE/KNC/440 from the ISRO-IISc Space Technology Cell.} \thanks{R.~G.~Gavaskar, C.~D.~Athalye, and K.~N.~Chaudhury are with the Department of Electrical Engineering, Indian Institute of Science, Bengaluru, India (e-mail: ruturajg@iisc.ac.in, chirayu@iisc.ac.in, kunal@iisc.ac.in).}}

\maketitle

\begin{abstract}
In plug-and-play (PnP) regularization, the knowledge of the forward model is combined with a powerful denoiser to obtain state-of-the-art image reconstructions. 
This is typically done by taking a proximal algorithm such as FISTA or ADMM, and formally replacing the proximal map associated with a regularizer by nonlocal means, BM3D or a CNN denoiser. 
Each iterate of the resulting PnP algorithm involves some kind of inversion of the forward model followed by denoiser-induced regularization. 
A natural question in this regard is that of \textit{optimality}, namely, do the PnP iterations minimize some $f+g$, where $f$ is a loss function associated with the forward model and $g$ is a regularizer? 
This has a straightforward solution if the denoiser can be expressed as a proximal map, as was shown to be the case for a class of linear symmetric denoisers. 
However, this result excludes kernel denoisers such as nonlocal means that are inherently non-symmetric. 
In this paper, we prove that a broader class of linear denoisers (including symmetric denoisers and kernel denoisers) can be expressed as a proximal map of some convex regularizer $g$. 
An algorithmic implication of this result  for non-symmetric denoisers is that it necessitates appropriate modifications in the PnP updates to ensure convergence to a minimum of $f+g$. 
Apart from the convergence guarantee, the modified PnP algorithms are shown to produce good restorations.
\end{abstract}

\begin{IEEEkeywords}
image reconstruction, plug-and-play regularization, linear denoiser, proximal map, convergence.
\end{IEEEkeywords}

\section{Introduction}
\label{sec:int}
\IEEEPARstart{I}{n computational} imaging, we are faced with the problem of reconstructing an unknown image $\boldsymbol{\xi} \in \Re^n$ (ground-truth) from partial and noisy 
measurements $\b \in \Re^m$ \cite{Gunturk2012_img_restoration,ribes2008linear,scherzer2009variational}. 
This includes restoration tasks such as denoising, inpainting, deblurring, superresolution \cite{scherzer2009variational,Park2003_superresolution}, and reconstruction modalities such as electron tomography, magnetic resonance imaging, single-photon imaging, inverse scattering, etc. \cite{Ercius2015_electron_tomography,Colton2013_scattering,Elgendy2016_single_photon}. 
The forward model in many of the above applications has the form $\b = \A \boldsymbol{\xi} + \boldsymbol{w} $, where $\A \in \Re^{m \times n}$ represents linear measurements and $\boldsymbol{w}$ is white Gaussian  noise \cite{ribes2008linear}.

The standard regularization framework for image reconstruction involves a data-fidelity (loss) function $f \colon \Re^n \to \Re$ and a regularizer $g \colon \Re^n \to \Re$; $f$ is derived from the forward model, while $g$ is typically derived from Bayesian or sparsity-promoting priors \cite{scherzer2009variational}. 
In this setup, the reconstruction is given by the solution of the following optimization problem:
\begin{equation}
\label{eq:main_prob}
\minimize_{\x \in \Re^n}\; f(\x) + g(\x).
\end{equation}
It is well-known that \eqref{eq:main_prob} corresponds to the maximum-a-posteriori (MAP) estimate of $\boldsymbol{\xi} $ given $\b$, where the likelihood of $\b$ is proportional to $\exp(-f(\x))$ and the prior distribution on $\x$ is proportional to $\exp(-g(\x))$. 
The regularizer $g$ plays the role of a prior (or bias) on the class of natural images and in effect forces the reconstruction to resemble the ground-truth \cite{Benning2018_regularization}. 
There has been extensive research on the design of image regularizers and efficient numerical solvers for \eqref{eq:main_prob}.
Some popular regularizers include sparsity-promoting functions such as total variation (TV) and wavelet regularizers \cite{Rudin1992_TV,Chambolle2010_TV_intro,Figueiredo2003_ISTA}, and low-rank regularizers such as weighted nuclear and Schatten norms of local image features \cite{Gu2014_WNNM,Gu2017_WNNM,Xie2016_weighted_schatten_min,Lefkimmiatis2013_hessian_schatten_norm}.
The regularizes in \cite{Gu2014_WNNM,Gu2017_WNNM,Xie2016_weighted_schatten_min} are however nonconvex; we solely focus on convex regularizers in this work.

\subsection{Plug-and-Play Regularization}
In the past few years, the plug-and-play (PnP) method \cite{Sreehari2016_PnP,Zhang2017_IRCNN,Dong2018_DNN_prior,Kamilov2017_PnP_ISTA,Ryu2019_PnP_trained_conv} has caught the attention of the imaging community, whereby powerful denoisers are leveraged  for  image regularization. This is achieved by taking a proximal algorithm for solving \eqref{eq:main_prob} such as FISTA or ADMM \cite{Beck2009_FISTA,Parikh2014_prox} and replacing the proximal map $\prox_g \colon \Re^n \to \Re^n$ of $g$,  
\begin{equation}
\label{eq:std_prox}
\prox_g(\y) := \argmin_{\x \in \Re^n}\; \frac{1}{2} \|\x - \y \|_2^2 + g(\x),
\end{equation}
with a Gaussian denoiser. The resulting algorithms are called PnP-FISTA, PnP-ADMM, etc., depending on the parent proximal algorithm. PnP is motivated by the connection between Gaussian denoising and the proximal map; in particular, note that \eqref{eq:std_prox} amounts to the MAP denoising of $\y$ given $\y = \boldsymbol{\xi}  +\boldsymbol{w}$, where $\w$ is white Gaussian noise \cite{Venkatakrishnan2013_PnP,Benning2018_regularization}. The idea behind PnP is to leverage superior denoisers such as nonlocal means (NLM) \cite{Buades2005_NLM}, BM3D \cite{Dabov2007_BM3D}, and DnCNN \cite{Zhang2017_DnCNN} for regularization, 
despite the limitation that they cannot be conceived as MAP estimators. 
PnP algorithms have been shown to produce state-of-the-art results in several imaging applications including  tomography, deblurring, superresolution, and compressive imaging \cite{Sreehari2016_PnP,Zhang2017_IRCNN,Dong2018_DNN_prior,Sun2019_PnP_SGD,Ryu2019_PnP_trained_conv,rick2017one,xiao2018discriminative,zhang2019deep,Kamilov2017_PnP_ISTA, Tirer2019_iter_denoising,Teodoro2019_PnP_fusion}.

\subsection{Motivation}
While PnP regularization works surprisingly well in practice, it is not a priori clear why PnP iterations should converge at all and whether the final reconstruction is optimal in some sense.
Indeed, it is not obvious if sophisticated denoisers such as NLM, BM3D, or DnCNN can be expressed as a proximal map. In fact, PnP-ADMM can in principle fail to converge even for a simple linear denoiser; see Appendix-\ref{subsec:counterexample}. Understanding the optimality and theoretical convergence of PnP regularization is thus an active research area \cite{Sreehari2016_PnP,Chan2017_PnP_conv,Dong2018_DNN_prior,Teodoro2019_PnP_fusion,Sun2019_PnP_SGD,Ryu2019_PnP_trained_conv,Tirer2019_iter_denoising,Gavaskar2020_PnP_ISTA,Xu2020_PnP_MMSE}. 
The question of optimality, and subsequently convergence, was positively resolved for a class of linear symmetric denoisers in  \cite{Sreehari2016_PnP,Teodoro2019_PnP_fusion}.
Till date, these are the only results on the optimality of PnP algorithms.
However, these results do not apply to denoisers such as kernel denoisers, e.g. NLM \cite{Milanfar2013_filtering_tour}, that are naturally non-symmetric; it is possible to symmetrize them \cite{Sreehari2016_PnP,Milanfar2013_filtering_tour}, but this comes with an overhead cost. This motivates the following questions:
\begin{enumerate}[(i)]
\item Can non-symmetric linear denoisers be expressed as the proximal map of some regularizer $g$?
\item If so, can we guarantee convergence of the PnP iterates to the minimum of $f+g$? 
\end{enumerate}
To the best of our knowledge, these are open questions and have not been addressed in the literature. A positive resolution of these questions would be of interest since kernel denoisers such as NLM are known to perform well for PnP regularization \cite{Gavaskar2020_PnP_ISTA,Unni2018_PnP_LADMM,Nair2019_HS_fusion,Sreehari2016_PnP}. Furthermore, as discussed in Section \ref{sec:linear_den}, there are a number of denoisers that are either exactly or approximately linear but not necessarily symmetric.  

\subsection{Contributions}
We investigate the above questions in this work. Our main results are summarized below:
\begin{enumerate}[(i)]
\item In Theorem \ref{thm:main}, we prove that under some conditions, a non-symmetric denoiser can be expressed as the proximal map of a convex regularizer $g$, but not in the standard form in \eqref{eq:std_prox}. More precisely, we can express the denoiser as a \textit{scaled} proximal map (Definition \ref{def:scaled-prox}), where the notion of \textit{proximity} arises from a weighted Euclidean distance.
Theorem \ref{thm:main} also subsumes an earlier result on symmetric linear denoisers; see \cite[Th. 2]{Teodoro2019_PnP_fusion}. Importantly, Theorem \ref{thm:main} applies to kernel denoisers which motivated the present work; see Section \ref{subsec:kernel_den} and Corollary \ref{cor:kernel}.
\item On the algorithmic side, in view of Theorem \ref{thm:main}, we show that it is necessary to introduce certain modifications in the PnP updates to ensure convergence to the minimum of $f+g$, where $g$ is the convex regularizer mentioned above. 
We make these changes in two commonly used PnP algorithms (PnP-FISTA and PnP-ADMM) and prove that the resulting iterates indeed converge to the minimum of $f+g$ if $f$ is closed, proper and convex (see Theorems \ref{thm:FISTA_scaled_conv} and \ref{thm:ADMM_scaled_conv}).
\end{enumerate}

\subsection{Organization}
The rest of the paper is organized as follows.
In Section \ref{sec:bg}, we present some background material and discuss prior work on PnP regularization.
In Section \ref{sec:linear_den}, we give an overview of linear denoisers and their use in PnP algorithms.
We state and discuss our main result about linear denoisers in Section \ref{sec:main}.
The proposed modification of PnP-FISTA and PnP-ADMM, and their convergence properties are discussed in Section \ref{sec:pnp}.
We numerically validate our convergence results in Section \ref{sec:exp}, and conclude with a summary of our findings in Section \ref{sec:conc}. 
Some technical proofs and discussions are deferred to the Appendix.

\subsection{Notation}
\label{subsec:notation}
The Euclidean norm is denoted as $\lVert \cdot \rVert_2$.
We use $\Sy^n_+$ (respectively, $\Sy^n_{++}$) to denote the set of $n \times n$ symmetric positive semidefinite (respectively, symmetric positive definite) matrices. 
For $\H \in \Sy^n_{++}$, we denote its symmetric square-root by $\H^{\frac{1}{2}}$.
If $\A \in \Re^{n \times n}$ has real eigenvalues $\lambda_1 \geq \cdots \geq \lambda_n$, we let $\lambda_{\max}(\mathbf{A})=\lambda_1$ and $\lambda_{\min}(\mathbf{A})=\lambda_n$.
We denote the identity matrix (of appropriate size) by $\I$.
We use $\mcR(\A)$ and $\mcN(\A)$ to denote the range space and the null space of $\A$.
We use $\Rc$ to denote the extended real line $[-\infty, \infty]$.
For $g \colon \Re^n \to \Rc$, the domain of $g$ is 
\begin{equation*}
\dom(g) := \{\x \in \Re^n : -\infty < g(\x) < \infty\}.
\end{equation*}
We denote the gradient, with respect to the Euclidean inner product, of a differentiable function $f \colon \Re^n \to \Re$ by $\nabla\!f$.
We use $i_C$ to denote the indicator function of a set $C \subseteq \Re^n$; more specifically, 
$i_C \colon \Re^n \to \Rc$ is given by 
\begin{equation}
\label{eq:indicator fn}
i_C(\x):= \left\{\begin{array}{rl} 0 & \mbox{ if } \x \in C  \\ \infty & \mbox{ if } \x \notin C. \end{array} \right.
\end{equation}

\section{Background}
\label{sec:bg}
\subsection{Optimization Preliminaries}
In this subsection, we succinctly recall some necessary preliminaries from optimization theory; see \cite{Parikh2014_prox,Rockafellar2009_variational_analysis,bubeck2015convex,Boyd2011_ADMM} for more details.
We denote the optimal value of \eqref{eq:main_prob} by $p^\star \in \Rc$; in other words,
\begin{equation*}
p^\star := \inf_{\x \in \Re^n}\; f(\x) + g(\x).
\end{equation*}
A point $\x^\star \in \Re^n$ is said to be a minimizer of \eqref{eq:main_prob} if $p^\star$ is finite and $f(\x^\star) + g(\x^\star) = p^\star$.
The optimization problem given by \eqref{eq:main_prob} can be equivalently written as follows:
\begin{mini}
{}{f(\x) + g(\z)}{\label{eq:main_prob_const}}{}
\addConstraint{\x}{=\z;}{}
\end{mini}
note that the optimal value of \eqref{eq:main_prob_const} is also $p^\star$.
The {\em Lagrangian}, $\mcL \colon (\Re^n \times \Re^n) \times \Re^n \to \Rc$, associated with \eqref{eq:main_prob_const} is given by
\begin{equation}
\label{eq:Lagrangian}
\mcL(\x,\z,\bnu) := f(\x) + g(\z) + \bnu^\top (\x - \z).
\end{equation}
A point $\big((\x^\star,\z^\star),\bnu^\star\big)$ is said to be a {\em saddle point} of the Lagrangian $\mcL$ if for all $\big((\x,\z),\bnu\big) \in (\Re^n \times \Re^n) \times \Re^n$,
\begin{equation*}
\mcL(\x^\star,\z^\star,\bnu) \leq \mcL(\x^\star,\z^\star,\bnu^\star) \leq \mcL(\x,\z,\bnu^\star).
\end{equation*}
Note that if $\big((\x^\star,\z^\star),\bnu^\star\big)$ is a saddle point of the Lagrangian given by \eqref{eq:Lagrangian}, then $(\x^\star,\z^\star)$ is a minimizer of \eqref{eq:main_prob_const}.
\begin{definition}
A function $g \colon \Re^n \to \Rc$ is said to be closed, proper and convex if its epigraph, 
\begin{equation*}
\epi(g) := \big\{(\x,t) \in \Re^n \times \Re : g(\x) \leq t \big\} \subseteq \Re^{n+1},
\end{equation*}
is a closed nonempty convex set and $g(\x) > -\infty, \forall \x \in \Re^n$.
\end{definition}
\begin{remark}
If $g \colon \Re^n \to \Rc$ is a closed proper convex function, then $\prox_g \colon \Re^n \to \Re^n$ given by \eqref{eq:std_prox} is well-defined; see e.g., \cite[Sec. 1.1]{Parikh2014_prox} and \cite[Th. 2.26]{Rockafellar2009_variational_analysis}. \fd
\end{remark}
\begin{definition}
\label{def:rho-smooth}
We say that $f \colon \Re^n \to \Re$ is $\rho$-smooth, where $\rho > 0$, if it is differentiable and
\begin{equation*}
\|\nabla\!f(\x) - \nabla\!f(\y)\|_2 \,\leq\, \rho\,  \|\x-\y\|_2 \qquad \forall \x,\y \in \Re^n.
\end{equation*}
\end{definition}
Note that if $f$ is $\rho$-smooth, then it is $\varepsilon$-smooth for all $\varepsilon \geq \rho$. 

\subsection{FISTA and ADMM}
We briefly discuss two commonly used proximal algorithms for solving \eqref{eq:main_prob}: FISTA (Fast Iterative Shrinkage-Thresholding
Algorithm) and ADMM (Alternating Direction Method of Multipliers); see \cite{Beck2009_FISTA,Chambolle2015_FISTA_conv,Boyd2011_ADMM,Eckstein1992_DouglasRachford,Chen2017_ADMM_note} for more details.

The FISTA update $(\y_{k},\x_{k})  \rightarrow (\y_{k+1},\x_{k+1})$, where $k = 1,2,\ldots,$ for \eqref{eq:main_prob} is as follows:  
\begin{subequations}
\label{eq:FISTA-itr}
\begin{align}
t_{k+1} &= \frac{1}{2} \left( 1 + \sqrt{1 + 4 t_k^2} \right), \label{eq:FISTA_t}\\
\y_{k+1} &= \x_k + \frac{t_k - 1}{t_{k+1}} (\x_k - \x_{k-1}), \label{eq:FISTA_y}\\
\x_{k+1} &= \prox_{g/\rho} \big( \y_{k+1} - \rho^{-1} \nabla\!f(\y_{k+1}) \big), \label{eq:FISTA_x}
\end{align}
\end{subequations}
where $t_1 = 1$, $\x_1 = \prox_{g/\rho} \big( \x_0 - \rho^{-1} \nabla\!f(\x_0) \big)$, $\x_0$ is an arbitrary initial point, and $\rho > 0$ is the {\em step-size}.
We state a standard result on FISTA (see \cite[Sec. 2 and Th. 4.4]{Beck2009_FISTA}) for ease of reference later in this paper.
\begin{theorem}[\cite{Beck2009_FISTA}]
\label{thm:FISTA_conv}
Suppose the following assumptions hold:
\begin{enumerate}[(i)]
\item $f \colon\Re^n \to \Re$ is convex and $\rho$-smooth.
\item $g \colon\Re^n \to \Rc$ is of the form $g = h + i_C$, where $h\colon \Re^n \to \Re$ is convex, $C$ is a nonempty closed convex set, and $i_C$ is given by \eqref{eq:indicator fn}.
\item Problem \eqref{eq:main_prob} has a minimizer $\x^\star$.
\end{enumerate}
Then the FISTA iterates \eqref{eq:FISTA_t}\,-\,\eqref{eq:FISTA_x} satisfy the following:
\begin{equation*}
f(\x_k) + g(\x_k) \leq p^\star + O(1/k^2).
\end{equation*}
\end{theorem}
\begin{remark}
\label{rmk:FISTA_iterate_conv}
If the $t$-update step \eqref{eq:FISTA_t} is altered to $t_{k+1} = 1+k/a$, where $a >2$, then as explained in \cite[Sec. 3]{Chambolle2015_FISTA_conv}, the iterates of this altered FISTA also satisfy the inequality in Theorem \ref{thm:FISTA_conv}. Moreover, by \cite[Th. 4.1]{Chambolle2015_FISTA_conv}, the corresponding sequence $(\x_k)_{k \in \Na}$ converges to a minimizer of \eqref{eq:main_prob}.  \fd
\end{remark}

The ADMM update $(\x_{k},\z_k,\bnu_k) \rightarrow (\x_{k+1},\z_{k+1},\bnu_{k+1})$, where $k = 1,2,\ldots,$ for \eqref{eq:main_prob_const} is as follows:   
\begin{subequations}
\label{eq:ADMM-itr}
\begin{align}
\x_{k+1} &= \prox_{f/\rho}(\z_k - \rho^{-1}\bnu_k), \label{eq:ADMM_prox_x}\\
\z_{k+1} &= \prox_{g/\rho}(\x_{k+1} + \rho^{-1}\bnu_k), \label{eq:ADMM_prox_z}\\
\bnu_{k+1} &= \bnu_k + \rho (\x_{k+1} - \z_{k+1}), \label{eq:ADMM_prox_nu}
\end{align}
\end{subequations}
where $\z_1,\bnu_1$ are arbitrary initial points, and $\rho > 0$ is the {\em penalty parameter}. 
The following theorem combines and restates \cite[Th. 8]{Eckstein1992_DouglasRachford} and \cite[Th. 4.1]{Chen2017_ADMM_note}; it is a standard result on the convergence of ADMM for problems of the form \eqref{eq:main_prob_const}. 
\begin{theorem}[\cite{Eckstein1992_DouglasRachford,Chen2017_ADMM_note}]
\label{thm:ADMM_conv}
Suppose $f$ and $g$ in \eqref{eq:main_prob_const} are closed proper convex functions, and the Lagrangian given by \eqref{eq:Lagrangian} has a saddle point.
Then for arbitrary $\rho > 0$ and $\z_1, \bnu_1 \in \Re^n$, the ADMM iterates \eqref{eq:ADMM_prox_x}\,-\,\eqref{eq:ADMM_prox_nu} converge to a saddle point of the Lagrangian; furthermore, $\lim_{k \to \infty} \big(f(\x_k) + g(\z_k)\big) = p^\star$.
\end{theorem}

PnP-FISTA and PnP-ADMM are obtained by formally replacing $\prox_{g/\rho}$  in  FISTA and ADMM (steps \eqref{eq:FISTA_x} and \eqref{eq:ADMM_prox_z})  with a denoiser $\mcD$. PnP-FISTA was first introduced in \cite{Kamilov2017_PnP_ISTA}, while PnP-ADMM was introduced in \cite{Venkatakrishnan2013_PnP}.
Note that PnP-FISTA is simpler to implement; however, it requires $f$ to be $\rho$-smooth.
On the other hand, PnP-ADMM is more powerful and it allows $f$ to be non-smooth; however, the $\x$-update step given by \eqref{eq:ADMM_prox_x} can be expensive to implement.

\subsection{Prior Work on PnP Regularization}
As mentioned earlier, it is not known if complex  denoisers such as BM3D and DnCNN can be expressed as a proximal map. 
The optimality question thus  remains open for such denoisers.  
The next best is to understand sequential convergence of the PnP iterations \cite{Ryu2019_PnP_trained_conv,Chan2017_PnP_conv,Gavaskar2020_PnP_ISTA}.
In this regard, some assumptions are typically  imposed on the denoiser, e.g., boundedness \cite{Chan2017_PnP_conv}, Lipschitz continuity \cite{Ryu2019_PnP_trained_conv}, or linearity \cite{Gavaskar2020_PnP_ISTA}.
In \cite{Dong2018_DNN_prior,Sun2019_PnP_SGD,Tirer2019_iter_denoising}, the authors establish subsequence and residue convergence under some assumptions on the denoiser.

The optimality of PnP regularization was originally investigated in \cite{Sreehari2016_PnP} for a class of linear denoisers that are (among other things) symmetric positive semidefinite. Specifically, it was shown that each denoiser in this class is the proximal map of a closed proper convex regularizer $g$. Furthermore, based on existing results on proximal algorithms, the authors could prove that the PnP iterations converge to the minimum of $f+g$ (assuming $f$ to be convex). The existence of the convex regularizer $g$ was predicted in  \cite{Sreehari2016_PnP} based on Moreau's classic theorem for proximal maps \cite{Moreau1965}, and the exact formula for $g$ was later worked out in \cite{Teodoro2019_PnP_fusion}. On a related note, it was recently shown in \cite{Xu2020_PnP_MMSE} that DnCNN can be approximately expressed as a proximal map of a nonconvex regularizer. The optimality of PnP regularization for nonlinear denoisers remains open as such. Even for linear denoisers, it remains unclear if the results in \cite{Sreehari2016_PnP,Teodoro2019_PnP_fusion} can be extended to non-symmetric denoisers such as kernel denoisers \cite{Singer2009_diffusion_nonlocal,Milanfar2013_filtering_tour}. This is precisely the topic of investigation in this paper.

\subsection{Moreau's Theorem for Proximal Maps}
\label{subsec:moreau's thm}
In this subsection, for ease of reference later in Section \ref{sec:main}, we briefly discuss an important result in \cite{Moreau1965} about proximal maps in an abstract Hilbert space.
Let $\big(\Xr, \langle \cdot\,,\cdot\rangle_\Xr \big)$ be a finite-dimensional real Hilbert space, and  
$\|\x\|_\Xr := \sqrt{\langle \x,\x \rangle_\Xr}$ be the norm induced by the inner product. 
The proximal map of a closed proper convex function $\Phi \colon \Xr \to \Rc$ is given by
\begin{equation*}
\label{eq:gen_prox}
\prox_{\Phi,\|\cdot\|_\Xr}\,(\y) := \argmin_{\x \in \Xr}\, \frac{1}{2} \|\x-\y\|_\Xr^2 + \Phi(\x).
\end{equation*}
The subdifferential of a convex function $\Psi: \Xr \to \Re$  is given by
\begin{equation*}
\partial \Psi(\y) := \big\{\a \in \Xr \,:\, \Psi(\z) \geq \Psi(\y) + \langle \a,\z-\y \rangle_\Xr \quad \forall \z \in \Xr\big\}.
\end{equation*}
The following theorem \cite[Corollary 10.c]{Moreau1965} gives a necessary and sufficient condition for $h \colon \Xr \to \Xr$ to be a proximal map.
\begin{theorem}[\cite{Moreau1965}]
\label{thm:moreau's thm}
A function $h \colon \Xr \to \Xr$ is the proximal map of some closed proper convex function $\Phi \colon \Xr \to \Rc$ if and only if the following conditions are satisfied:
\begin{enumerate}[(i)]
\item There exists a convex function $\Psi \colon \Xr \to \Re$ such that $h(\y) \in \partial \Psi(\y)$ for all $\y \in \Xr$.
\item $h$ is non-expansive, i.e., $\|h(\x)-h(\y)\|_\Xr \leq \|\x-\y\|_\Xr$ for all $\x,\y \in \Xr$.
\end{enumerate}
\end{theorem}

\section{Linear Denoisers in PnP Algorithms}
\label{sec:linear_den}
We begin with an overview of linear denoisers and their use in PnP algorithms.
Popular denoisers such as BM3D and DnCNN are typically nonlinear and difficult to analyze.
Understanding how linear denoisers behave with PnP algorithms is a first step towards unraveling sophisticated nonlinear denoisers.
As discussed next, there are denoisers that are exactly or approximately linear, and they perform quite well in practice.

A linear denoiser $\mcD : \Re^n \to \Re^n$ is one that can be expressed as a linear transform, i.e., given an input image $\x \in \Re^n$,
\begin{equation}
\label{eq:def-mcD}
\mcD(\x) = \W\x
\end{equation}
for some $\W \in \Re^{n \times n}$.
Examples include the box filter, the Gaussian filter, and the Wiener filter. These classical filters denoise an image by replacing each pixel by a weighted average of its neighbors \cite{Gunturk2012_img_restoration}.
For more recent examples of linear denoisers, see \cite{Singer2009_diffusion_nonlocal,Haque2015_symmetric_filters,Wei2015_sparse_fusion,Teodoro2019_PnP_fusion}.
We next discuss a special family of denoisers called kernel denoisers.

\subsection{Kernel Denoisers}
\label{subsec:kernel_den}
In the box and Gaussian filters, $\W$ is fixed (up to a choice of parameters) and does not depend on the input.
To improve the denoising performance, the weights can be adapted from pixel to pixel.
Indeed, in modern denoisers such as NLM, the weights are derived from the input image \cite{Buades2005_NLM,Singer2009_diffusion_nonlocal,Milanfar2013_filtering_tour}.
As far as the input-output relationship is concerned, such a denoiser is no longer linear; rather, it can be represented as $\mcD(\x) = \W \x$, where $\W$ depends on $\x$.
For denoising purpose, the components of $\W$ should be nonnegative, and the sum of every row of $\W$ should be $1$.
One way to achieve this is as follows.
We choose a kernel function $\kappa$ that measures the affinity between two pixels using some local features.
The pairwise kernel values (usually a subset of pairwise pixels are used) are collected in a kernel matrix $\K \in \Re^{n \times n}$.
Typically, $\kappa$ is chosen so that $\K$ has the following properties \cite{Singer2009_diffusion_nonlocal,Milanfar2013_filtering_tour}:
\begin{enumerate}[(i)]
\item $\K$ is nonnegative and symmetric positive semidefinite.
\item The row-sums of $\K$ are positive.
\end{enumerate}
The row-sums are collected in a diagonal matrix  $\D$ (we call $\D$ the \textit{normalization matrix}) and the weight matrix is given by
\begin{equation}
\label{kernelfilter}
\W= \D^{-1} \K.
\end{equation}
Though $\K \in \Sy^n_+$ and $\D \in \Sy^n_{++}$, note that $\W$ given by \eqref{kernelfilter} is generally non-symmetric.
We refer to denoisers of the form \eqref{kernelfilter} as {\em kernel denoisers}; this includes the bilateral filter, NLM, and LARK \cite{Milanfar2013_filtering_tour}.

Since $\W$ depends on $\x$, a kernel denoiser is nonlinear. 
However, by effectively removing the dependence of $\W$ on $\x$, we can turn it into a linear filter. 
One approach is to use a surrogate image $\y$ (e.g., a prefiltered version of $\x$) to compute $\W$, whereby $\mcD(\x) = \W(\y) \x$ becomes linear in $\x$ \cite{Milanfar2013_filtering_tour}. 
A different approach  \cite{Sreehari2016_PnP} that is specifically geared towards PnP algorithms is as follows. We treat $\W$ as a kernel filter in the first $j \sim 10$ iterations, and for the rest of the iterations, we fix $\W$ to be the weight matrix at the end of the $j$-th iteration. 
This allows one to exploit the adaptability of kernel filters in the first $j$ iterations, before turning it into a linear denoiser for the remaining iterations. 
An important point to note here is that the filter is linear except for a finite number of iterations, which can be ignored in the convergence analysis.
This approach was adopted in \cite{Gavaskar2020_PnP_ISTA,Unni2018_PnP_LADMM} to analyze PnP convergence.
In this paper, we consider kernel denoisers to be linear in the above sense.
Henceforth, for notational convenience, we use $\W$ to denote a linear denoiser given by \eqref{eq:def-mcD}.

A closely related class of linear denoisers is the so-called $2\W - \W^2$ filters \cite{Singer2009_diffusion_nonlocal}.
As the name suggests, a denoiser in this class is of the form $\mcD(\x) = (2\W - \W^2)\x$, where $\W = \D^{-1} \K$ is a kernel denoiser.
As discussed in \cite{Singer2009_diffusion_nonlocal}, the function $2\W - \W^2$ exhibits good denoising properties, although its edge-preserving properties are somewhat different from those of $\W$.

\subsection{PnP using Symmetric Denoisers}
If $\W \in \Sy_{+}^n$ and $\lambdamax(\W) \leq 1$, it is known that $\W$ is the proximal map of a closed proper convex regularizer \cite{Sreehari2016_PnP,Teodoro2019_PnP_fusion}.
The precise result \cite[Th. 2]{Teodoro2019_PnP_fusion} is stated next.
Note that if $\W \in \Sy^n_+$ and $\rank(\W) = r$, then it has a condensed eigenvalue decomposition $\W = \V_r \LA_r \V^\top_r$, where $\LA_r \in \Re^{r \times r}$ is a diagonal matrix containing the positive eigenvalues of $\W$, and $\V_r \in \Re^{n \times r}$ is the matrix whose columns are the corresponding orthonormal eigenvectors.
\begin{theorem}[\cite{Teodoro2019_PnP_fusion}]
\label{thm:teodoro}
Suppose $\W \in \Sy^n_+$ and $\lambdamax(\W) \in (0,1]$.
Let $\V_r \LA_r \V^\top_r$ be a condensed eigenvalue decomposition of $\W$.
Then $\W \colon \Re^n \to \Re^n$ is the proximal map of a closed proper convex function $\Phi \colon \Re^n \to \Rc$ given by
\begin{equation}
\label{eq:phi_teodoro}
\Phi(\x) := i_{\mcR(\W)}(\x) + \frac{1}{2} \x^\top \V_r (\LA^{-1}_r - \I) \V^\top_r \x,
\end{equation}
where $i_{\mcR(\W)}$ is the indicator function of $\mcR(\W) \subseteq \Re^n$.
\end{theorem}

Examples of denoisers that satisfy the conditions in Theorem \ref{thm:teodoro} are the GMM-based denoiser in \cite{Teodoro2019_PnP_fusion,Sulam2016_GM_fusion}, the dictionary-based denoiser in \cite{Wei2015_sparse_fusion}, and the DSG-NLM denoiser \cite{Sreehari2016_PnP}; the latter is a symmetric approximation of NLM.
However, since $\W$ is generally non-symmetric for kernel denoisers, Theorem \ref{thm:teodoro} does not subsume kernel denoisers.
\begin{remark}
\label{rmk:divergence}
For PnP-ADMM using a symmetric $\W$ that satisfies the conditions in Theorem \ref{thm:teodoro}, the convergence properties in Theorem \ref{thm:ADMM_conv} hold. A natural question is whether we can expect this result to hold under similar conditions for  non-symmetric $\W$. A counterexample is provided in Appendix-\ref{subsec:counterexample}; we construct a non-symmetric $\W$ whose eigenvalues are nonnegative and $\lambdamax(\W) \in (0,1]$, but PnP-ADMM does not exhibit the convergence in Theorem \ref{thm:ADMM_conv}. \fd
\end{remark}

\section{Which Linear Denoisers are Proximal Maps?}
\label{sec:main}
In this section, we state our main result on linear denoisers, which gives a different characterization for a linear denoiser to be a {\em scaled} proximal map.

\subsection{Scaled Proximal Maps}
We briefly discuss below the concept of scaled proximal maps; see \cite{Friedlander2017_efficient_scaled_prox} and \cite{Park2019_VM_ISTA} for more details.
Note that given $\H \in \Sy^n_{++}$, we can associate the following inner product and norm on $\Re^n$: 
\begin{eqnarray}
\label{eq: H-inner-product}
\langle \x,\y\rangle_{\H} := \x^\top \H \y &\mbox{and}& \normH{\x} := \sqrt{\x^\top \H \x}. 
\end{eqnarray}
\begin{definition}
\label{def:scaled-prox}
Let $g \colon\Re^n \to \Rc$ be closed, proper and convex.
The scaled proximal map of $g$, with respect to $\H \in \Sy^n_{++}$, is defined as follows:
\begin{equation}
\label{eq:scaled_prox}
\prox_{g,\normH{\cdot}}\,(\y) := \argmin_{\x \in \Re^n}\, \frac{1}{2} \normH{\x-\y}^2 + g(\x) .
\end{equation}
\end{definition}
We refer to this as the {\em $\H$-scaled proximal map}  of $g$, where $\H \in \Sy^n_{++}$ is called the {\em scaling matrix}.
Note that when $\H = \I$, we recover the standard proximal map of $g$, which we henceforth denote by $\prox_{g,\|\cdot\|_2}$.
\begin{remark}
\label{rmk:scaled vs general prox-maps}
The notion of the $\H$-scaled proximal map is the same as that of the proximal map in the Hilbert space $\big(\Re^n, \langle \cdot\,,\cdot\rangle_\H \big)$, as discussed in Section \ref{subsec:moreau's thm}. \fd
\end{remark}

\subsection{Linear Denoisers as Scaled Proximal Maps}
The following theorem is one of the main results of this paper. It gives a necessary and sufficient condition for a linear denoiser to be the scaled proximal map of a closed proper convex function having some specific form.
\begin{theorem}
\label{thm:main} The following statements are equivalent.
\begin{enumerate}[(i)]
\item $\W$ is the scaled proximal map, w.r.t. some $\H \in \Sy_{++}^n$, of a closed proper convex function $\Phi \colon \Re^n \to \Rc$ having the following form (up to a constant):
\begin{equation}
\label{eq:phi}
\Phi(\x) := i_{\mcR(\W)}(\x) + \frac{1}{2} \x^\top \P \x, \qquad \big(\P \in \Sy^n_+\big).
\end{equation}
\item $\W$ is similar to some $\M \in \Sy^n_+$ and $\lambdamax(\W) \leq 1$.
\end{enumerate}
\end{theorem}
\begin{IEEEproof}
(ii) $\Rightarrow$ (i): As $\W$ is similar to some $\M \in \Sy^n_+$ and $\lambdamax(\W) \leq 1$, we have $\lambdamax(\W) \in [0,\,1]$. 
If $\lambdamax(\W) = 0$, then $\W = \ZE$. 
Note that $\W = \ZE$ is the standard proximal map of a closed proper convex function $\Phi \colon \Re^n \to \Rc$ of the form given by \eqref{eq:phi}, where $\P \in \Sy^n_+$ can be arbitrary. 
Next, we consider the case where $\lambdamax(\W) \in (0,\,1]$.

Let $\rank(\W) = r$.
Since $\W$ is similar to some $\M \in \Sy^n_+$ and $\lambdamax(\W) \in (0,\,1]$, it has an eigenvalue decomposition  given by $\W = \V \LA \V^{-1}$, where $\LA \in \Sy^n_+$ is a diagonal matrix whose first $r$ diagonal entries are in $(0,1]$ and the remaining entries are $0$.
Let $\LA_r \in \Sy_{++}^r$ be the $(r \times r)$ principal submatrix of $\LA$.
Let $\U \in \Re^{n \times r}$ and $\U^\dagger \in \Re^{r \times n}$ be the matrices consisting of the first $r$ columns of $\V$ and the first $r$ rows of $\V^{-1}$, respectively.
Thus, $\U \LA_r \U^\dagger$ is a condensed eigenvalue decomposition of $\W$.
Let $\Phi \colon \Re^n \rightarrow \Rc$ be given by
\begin{equation}
\label{eq:phi_proof}
\Phi(\x) := i_{\mcR(\W)}(\x) + \frac{1}{2} \ \x^\top (\U^\dagger)^\top (\LA_r^{-1} - \I) \U^\dagger\x.
\end{equation}
Since the diagonal entries of $\LA_r$ belong to $(0,1]$, we have $(\LA_r^{-1} - \I) \in \Sy^r_+$; and hence, $(\U^\dagger)^\top (\LA_r^{-1} - \I)\U^\dagger \in \Sy^n_+$.
Thus, $\Phi$ is a closed proper convex function of the form \eqref{eq:phi}.

Let $\H := (\V \V^\top)^{-1} \in \Sy^n_{++}$.
It follows from Definition \ref{def:scaled-prox} and \eqref{eq:phi_proof} that for all $\y \in \Re^n$,
\begin{equation}
\label{eq:scaled_prox_op_phi eq-1}
\prox_{\Phi,\normH{\cdot}}(\y) = \argmin_{\x \in \mcR(\W)}\, \normH{\x - \y}^2 + \x^\top \P \x,
\end{equation}
where $\P := (\U^\dagger)^\top (\LA_r^{-1} - \I) \U^\dagger$. 
Note that for an arbitrary $\x \in \mcR(\W)$, there exists a unique $\z \in \Re^r$ such that $\x = \U \z$.
Consequently, using the fact that  $\U^\dagger \U = \I \in \Re^{r \times r}$, we can rewrite \eqref{eq:scaled_prox_op_phi eq-1} as follows:
\begin{equation}
\label{eq:scaled_prox_op_phi eq-2}
\prox_{\Phi,\normH{\cdot}}(\y) = \U \left( \argmin_{\z \in \Re^r}\; \alpha_{\y}(\z) \right) \qquad \forall \y \in \Re^n,
\end{equation}
where $\alpha_{\y}(\z) := \normH{\U\z - \y}^2 + \z^\top (\LA_r^{-1} - \I) \z$.

As $\H = (\V \V^\top)^{-1}$, using the definitions of $\U$ and $\U^\dagger$, we can conclude that $\U^\dagger = \U^\top \H$; thus, $\U^\top \H \U = \I$. 
Therefore, for an arbitrary $\y \in \Re^n$, we have
\begin{equation}
\label{eq:exprsn alpha_y}
\alpha_{\y}(\z) = \z^\top \LA_r^{-1} \z - 2\z^\top \U^\dagger \y + \y^\top \H \y \qquad \forall \z \in \Re^r.
\end{equation}
As $\alpha_{\y} \colon \Re^r \rightarrow \Re$ is a strictly convex quadratic function, it follows from \eqref{eq:scaled_prox_op_phi eq-2} and \eqref{eq:exprsn alpha_y} that
\begin{equation*}
\prox_{\Phi,\normH{\cdot}}(\y) = \U \big(\LA_r \U^\dagger\y\big) = \W \y \qquad \forall \y \in \Re^n. 
\end{equation*}

(i) $\Rightarrow$ (ii): $\W \in \Re^{n \times n}$ is the scaled proximal map, w.r.t some $\H \in \Sy^n_{++}$, of a closed proper convex function $\Phi \colon \Re^n \rightarrow \Rc$ of the form given by \eqref{eq:phi}; i.e., for all $\y \in \Re^n$,
\begin{equation}
\label{eq:scaled_prox_op_phi eq-i}
\W \y = \argmin_{\x \in \mcR(\W)}\, \normH{\x-\y}^2 + \x^\top \P \x.
\end{equation}
If $\W = \ZE$, then it obviously satisfies the given condition. We next consider the case $\W \neq \ZE$. Let $\rank(\W) = r$ and $\Q \in \Re^{n \times r}$ be a matrix whose columns form a basis of $\mcR(\W)$. 
For an arbitrary $\x \in \mcR(\W)$, there exists a unique $\z \in \Re^r$ such that $\x = \Q \z$.
Thus, we can rewrite \eqref{eq:scaled_prox_op_phi eq-i} as follows:
\begin{equation}
\label{eq:scaled_prox_op_phi eq-ii}
\W \y = \Q \left( \argmin_{\z \in \Re^r}\; \beta_{\y}(\z)\right) \qquad \forall \y \in \Re^n,
\end{equation}
where $\beta_{\y}(\z) := \z^\top \Q^\top (\H+\P) \Q \z - 2\z^\top \Q^\top \H \y + \y^\top \H \y$.
Let $\R := \Q^\top (\H+\P) \Q$; as $(\H+\P) \in \Sy^n_{++}$ and $\Q \in \Re^{n \times r}$ has full column rank, we have $\R \in \Sy^r_{++}$. 
Thus, $\beta_{\y}$ is a strictly convex quadratic function.
Now, it follows from \eqref{eq:scaled_prox_op_phi eq-ii} and the definition of $\beta_{\y}$ that
\begin{equation*}
\W \y = \Q \big(\R^{-1} \Q^\top \H \y\big) \qquad \forall \y \in \Re^n.
\end{equation*}
Let $\S := \Q \R^{-1} \Q^\top \in \Sy^n_+$; thus, $\W = \S \H$.

Note that $\W = \S \H = \H^{-\frac{1}{2}} \big(\H^{\frac{1}{2}} \S \H^{\frac{1}{2}} \big) \H^{\frac{1}{2}}$.
Therefore, $\W$ is similar to $\H^{\frac{1}{2}} \S \H^{\frac{1}{2}} \in \Sy_{+}^n$. 
Furthermore, using Theorem \ref{thm:moreau's thm}, we can conclude that $\W$ is non-expansive with respect to  $\|\cdot\|_\H$. 
In other words, $\|\W\y\|_\H \leq \|\y\|_\H$ for all $\y \in \Re^n$.
Therefore, $\lambdamax(\W) \leq 1$.
\end{IEEEproof}
\begin{remark}
Theorem \ref{thm:main} can be viewed as a special case of Theorem \ref{thm:moreau's thm} for linear functions.
However, for this special case, unlike the existential statement in Theorem \ref{thm:moreau's thm}, we have provided the expression of a closed proper convex function $\Phi \colon \Re^n \to \Rc$ corresponding to a linear proximal map $\W$. \fd
\end{remark}

To better understand the connection between Theorem \ref{thm:main} and Theorem \ref{thm:moreau's thm}, consider $\Psi \colon \Re^n \to \Re$ given by
\begin{equation}
\label{eq:psi}
\Psi(\y) := \frac{1}{2} \ \y^\top \H \W \y = \frac{1}{2} \ \langle \y, \W\y\rangle_\H,
\end{equation}
where $\W$ and $\H$ are as given in Theorem \ref{thm:main}.
It follows from the proof of Theorem \ref{thm:main} that $\W \in \Re^{n \times n}$ can be written as $\W = \S \H$, where $\S \in \Sy^n_+$. 
Therefore, $\H \W = \H \S \H \in \Sy^n_+$; and hence, $\Psi$ given by \eqref{eq:psi} is a convex function.
Furthermore, note that for all $\x,\y \in \Re^n$,
\begin{equation}
\label{eq:grad_psi}
\nabla \Psi(\y)^\top \x = \y^\top \W^\top \H \x =  \langle \W \y, \x\rangle_\H.
\end{equation}
It follows from the first-order condition for convex functions and \eqref{eq:grad_psi} that for all $\y,\z \in \Re^n$, 
\begin{equation*}
\Psi(\z) \geq \Psi(\y) + \nabla \Psi(\y)^\top (\z-\y) = \Psi(\y) + \langle \W \y,\z-\y \rangle_\H.
\end{equation*}
Therefore, in the Hilbert space $\big(\Re^n, \langle \cdot\,,\cdot\rangle_\H \big)$, we can conclude that $\W \y \in \partial \Psi(\y)$; in fact, since $\Psi$ is differentiable, we have $\partial \Psi(\y) = \{\W \y\}$ for all $\y \in \Re^n$.
Thus, as explained in Remark \ref{rmk:scaled vs general prox-maps}, the appropriate Hilbert space in the context of Theorem \ref{thm:main} is $\big(\Re^n, \langle \cdot\,,\cdot\rangle_\H \big)$.
\begin{remark}
\label{rmk:proof_consequences}
It follows from the proof of Theorem \ref{thm:main} that:
\begin{enumerate}[(i)]
\item The pair $(\H,\P)$ in the statement of Theorem \ref{thm:main} can be replaced by $(c \H, c \P)$, where $c > 0$.
\item When $\W \in \Sy^n_{+}$, we can have $\H = \I$; consequently, Theorem \ref{thm:main} subsumes Theorem \ref{thm:teodoro}. \fd
\end{enumerate}
\end{remark}

Note that in the proof of Theorem \ref{thm:main}, we have explicitly constructed a scaling matrix $\H \in \Sy^n_{++}$ from $\W$; this has an interesting consequence for kernel denoisers.
Recall that (see Section \ref{subsec:kernel_den}) a kernel matrix $\K$ is nonnegative and symmetric positive semidefinite, and its row-sums are positive.
\begin{corollary}
\label{cor:kernel}
Let $\W = \D^{-1} \K$ be a kernel denoiser, where $\K \in \Sy^n_{+}$ is a kernel matrix and $\D \in \Sy_{++}^n$ is the corresponding normalization matrix.
Then $\W$ is the $\D$-scaled proximal map of a closed proper convex function.
\end{corollary}
\begin{IEEEproof}
Note that $\W = \D^{-\frac{1}{2}} \M \D^{\frac{1}{2}}$, where $\M := \D^{-\frac{1}{2}} \K \D^{-\frac{1}{2}} \in \Sy_{+}^n$.
By construction, $\|\W\|_{\infty} = 1$, where $\|\W\|_{\infty}$ is the induced $\infty$-norm (see \cite[Sec. 2.1]{Watkins2010_matrix_comp}) of $\W$.
Thus, $\lambdamax(\W) \leq \|\W\|_{\infty} = 1$; and hence $\W$ satisfies the equivalent condition given in Theorem \ref{thm:main}.
Now, let $\G$ be an orthogonal matrix of eigenvectors of $\M$.
Note that the columns of $\V := \D^{-\frac{1}{2}} \G$ are eigenvectors of $\W$.
It follows from the proof of Theorem \ref{thm:main} that $(\V \V^\top)^{-1} = \D$ is the desired scaling matrix $\H$ in Theorem \ref{thm:main}.
\end{IEEEproof}

Note that $2\W - \W^2$ filters, where $\W = \D^{-1} \K$, also satisfy the equivalent condition in Theorem \ref{thm:main}.
Moreover, the scaling matrix for such a filter is $\D$.

\section{Scaled PnP Algorithms}
\label{sec:pnp}
Consider a linear denoiser which satisfies the equivalent condition in Theorem \ref{thm:main}.
We validate in Appendix-\ref{subsec:counterexample} that for a standard PnP proximal algorithm using such a denoiser, the convergence properties of the parent proximal algorithm need not hold.
In this section, we resolve this issue by providing some modifications to PnP-FISTA and PnP-ADMM. We call these modified algorithms, which are coherent with the notion of scaled proximal maps, as {\em scaled} PnP-FISTA and {\em scaled} PnP-ADMM, respectively.
We show that the scaled PnP algorithms, when used with a linear denoiser $\W$ that is {\em proximable} in the sense of Theorem \ref{thm:main}, inherit the convergence properties of their parent proximal algorithms. 
Note that the discussion in this section has a natural counterpart for any PnP proximal algorithm; see among others \cite{Friedlander2017_efficient_scaled_prox,Gonccalves2017_VM_PADMM,Park2019_VM_ISTA} for a discussion on proximal algorithms based on scaled proximal maps. 

The update $(\y_{k},\x_{k}) \rightarrow (\y_{k+1},\x_{k+1})$, where $k = 1,2,\ldots,$ for scaled PnP-FISTA is defined as follows:  
\begin{subequations}
\begin{align}
t_{k+1} &= \frac{1}{2} \left( 1 + \sqrt{1 + 4 t_k^2} \right), \label{eq:FISTA_scaled_t}\\
\y_{k+1} &= \x_k + \frac{t_k - 1}{t_{k+1}} (\x_k - \x_{k-1}), \label{eq:FISTA_scaled_y}\\
\x_{k+1} &= \mcD \big(\y_{k+1} - \rho^{-1}\H^{-1} \nabla\!f(\y_{k+1})\big), \label{eq:FISTA_scaled_x}
\end{align}
\end{subequations}
where $t_1 = 1$, $\x_1 = \mcD \big(\x_0 - \rho^{-1}\H^{-1} \nabla\!f(\x_0)\big)$, $\x_0$ is arbitrary and $\H \in \Sy_{++}^n$.
We say $f \colon \Re^n \to \Re$ is $\rho$-smooth with respect to $\H \in \Sy^n_{++}$ if it is differentiable and for all $\x,\y \in \Re^n$,
\begin{equation}
\label{eq:rho-H-smooth}
\normH{\H^{-1} \nabla\!f(\x) - \H^{-1}\nabla\!f(\y)} \,\leq\, \rho \normH{\x-\y}.
\end{equation}
Note that when $\H=\I$, we get Definition \ref{def:rho-smooth}.
The following theorem asserts the objective convergence, under conditions similar to Theorem \ref{thm:FISTA_conv}, of the scaled PnP-FISTA algorithm which uses an appropriate linear denoiser.
\begin{theorem}
\label{thm:FISTA_scaled_conv}
Let $\mcD$ be a linear denoiser given by \eqref{eq:def-mcD}, where $\W$ satisfies the equivalent condition given in Theorem \ref{thm:main}; let $\Phi$ and $\H$ be as given in Theorem \ref{thm:main}.
Suppose the following conditions hold:
\begin{enumerate}[(i)]
\item $f \colon\Re^n \to \Re$ is convex and $\rho$-smooth w.r.t. $\H \in \Sy^n_{++}$.
\item Problem \eqref{eq:main_prob}, with $g := \rho \Phi$, has a minimizer $\x^\star$.
\end{enumerate}
Then the iterates \eqref{eq:FISTA_scaled_t}\,-\,\eqref{eq:FISTA_scaled_x} of scaled PnP-FISTA satisfy
\begin{equation}
\label{eq:conv1}
f(\x_k) + g(\x_k) \leq p^\star + O(1/k^2),
\end{equation}
where $p^\star$ is the optimal value of \eqref{eq:main_prob}.
\end{theorem}
\begin{IEEEproof}
See Appendix-\ref{subsec:proof_FISTA_scaled_conv}.
\end{IEEEproof}
\begin{remark}
It follows from Remark \ref{rmk:FISTA_iterate_conv} that if \eqref{eq:FISTA_scaled_t} is altered to $t_{k+1} = 1+k/a$, where $a >2$, then Theorem \ref{thm:FISTA_scaled_conv} can be strengthened to additionally conclude the convergence of the sequence $(\x_k)_{k \in \Na}$ to a minimizer of \eqref{eq:main_prob}; its proof follows on similar lines. \fd
\end{remark}
\begin{remark}
For a general linear denoiser satisfying the equivalent condition given in Theorem \ref{thm:main}, the $\x$-update step given by \eqref{eq:FISTA_scaled_x} can be numerically challenging as it involves computing $\H^{-1} \nabla\!f(\y_{k+1})$.
However, for kernel denoisers, where $\H = \D$, this simply reduces to elementwise division. \fd
\end{remark}

\begin{figure*}[t!]
\centering
\subfloat[\textit{Aerial.}]{\includegraphics[width=0.12\linewidth]{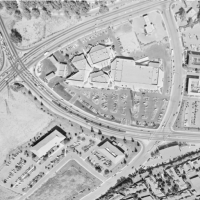}}
\hfill
\subfloat[\textit{Bridge.}]{\includegraphics[width=0.12\linewidth]{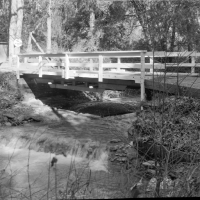}}
\hfill
\subfloat[\textit{Car.}]{\includegraphics[width=0.12\linewidth]{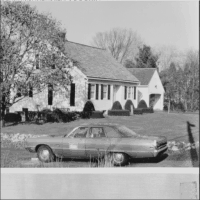}}
\hfill
\subfloat[\textit{Clock.}]{\includegraphics[width=0.12\linewidth]{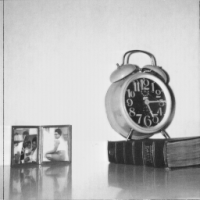}}
\hfill
\subfloat[\textit{Goldhill.}]{\includegraphics[width=0.12\linewidth]{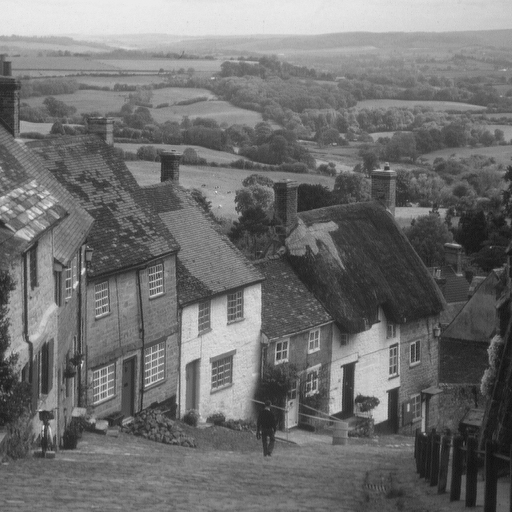}}
\hfill
\subfloat[\textit{Lake.}]{\includegraphics[width=0.12\linewidth]{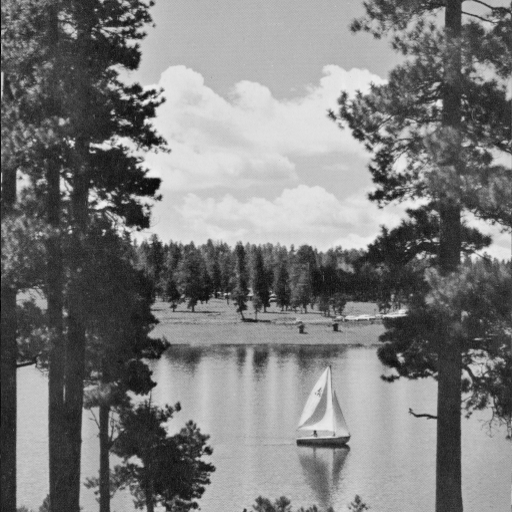}}
\hfill
\subfloat[\textit{Mandril.}]{\includegraphics[width=0.12\linewidth]{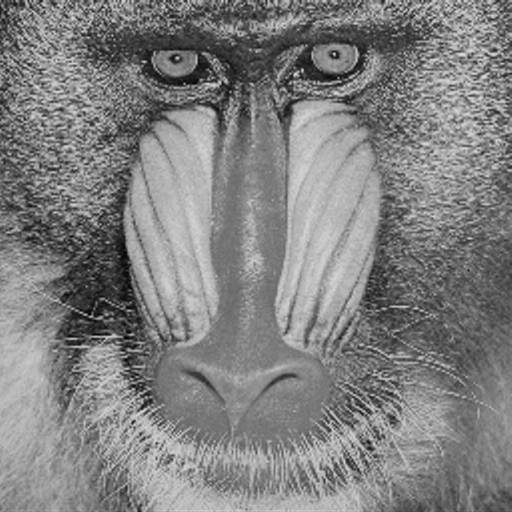}}
\hfill
\subfloat[\textit{Tree.}]{\includegraphics[width=0.12\linewidth]{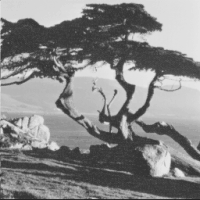}}
\caption{Eight test images (other than those in the Set12 dataset) that make up our twenty-image dataset.}
\label{fig:image_set}
\end{figure*}

The update $(\x_{k},\z_k,\bnu_k) \rightarrow (\x_{k+1},\z_{k+1},\bnu_{k+1})$, where $k = 1,2,\ldots,$ for scaled PnP-ADMM is defined as follows:
\begin{subequations}
\begin{align}
\x_{k+1} &= \prox_{f/\rho,\normH{\cdot}}\!(\z_k - \rho^{-1}\bnu_k), \label{eq:ADMM_scaled_prox_x}\\
\z_{k+1} &= \mcD(\x_{k+1} + \rho^{-1}\bnu_k), \label{eq:ADMM_scaled_prox_z}\\
\bnu_{k+1} &= \bnu_k + \rho(\x_{k+1} - \z_{k+1}) \label{eq:ADMM_scaled_prox_nu},
\end{align}
\end{subequations}
where $\z_1, \bnu_1$ are arbitrary initial points, $\rho > 0$ is the penalty parameter, and $\H \in \Sy_{++}^n$.
For $\H \in \Sy^n_{++}$, the $\H$-scaled Lagrangian (denoted by $\mcL_{\H}$) of the problem \eqref{eq:main_prob_const} is defined as follows (see \cite[Sec. 2]{Ito1990_ALM_hilbert}):
\begin{equation}
\label{eq:scaled_Lagrangian-main_prob_const}
\mcL_{\H}(\x,\z,\bnu) := f(\x) + g(\z) + \langle \bnu,\x-\z \rangle_{\H}.
\end{equation}
Similar to the case of standard Lagrangian, if $\big((\x^\star,\z^\star),\bnu^\star\big)$ is a saddle point of $\mcL_{\H}$, then $(\x^\star,\z^\star)$ is a minimizer of \eqref{eq:main_prob_const}.
The following theorem guarantees, under conditions similar to Theorem \ref{thm:ADMM_conv}, the convergence of the scaled PnP-ADMM algorithm which uses an appropriate linear denoiser.
\begin{theorem}
\label{thm:ADMM_scaled_conv}
Let $\mcD$ be a linear denoiser given by \eqref{eq:def-mcD}, where $\W$ satisfies the equivalent condition given in Theorem \ref{thm:main}; let $\Phi$ and $\H$ be as given in Theorem \ref{thm:main}. Let $\rho > 0$ be the penalty parameter in \eqref{eq:ADMM_scaled_prox_x}.
Suppose for the problem \eqref{eq:main_prob_const}, with $g := \rho \Phi$, the following conditions hold:
\begin{enumerate}[(i)]
\item $f$ is a closed proper convex function.
\item $\mcL_{\H}$ given by \eqref{eq:scaled_Lagrangian-main_prob_const} has a saddle point.
\end{enumerate}
Then for arbitrary $\z_1, \bnu_1 \in \Re^n$, the iterates \eqref{eq:ADMM_scaled_prox_x}\,-\,\eqref{eq:ADMM_scaled_prox_nu} of scaled PnP-ADMM converge to a saddle point of $\mcL_{\H}$; furthermore, $\lim_{k \to \infty} \big(f(\x_k) + g(\z_k)\big) = p^\star$.
\end{theorem}
\begin{IEEEproof}
See Appendix-\ref{subsec:proof_ADMM_scaled_conv}.
\end{IEEEproof}
\begin{remark}
The conditions in Theorem \ref{thm:ADMM_scaled_conv} are mild.
In particular, the saddle point condition, which is equivalent to strong duality, holds in many image restoration problems; we refer readers to \cite[Appendix A]{Sreehari2016_PnP} for a detailed discussion. \fd
\end{remark}

Note that when $\H = \I$, scaled PnP-FISTA (scaled PnP-ADMM) reduces to standard PnP-FISTA (standard PnP-ADMM).

\section{Numerical Results}
\label{sec:exp}
In this section, we numerically validate the convergence results in Theorems \ref{thm:FISTA_scaled_conv} and \ref{thm:ADMM_scaled_conv} for two restoration applications, image inpainting and image deblurring.
The forward model for both applications is given by $\b = \A \boldsymbol{\xi} + \w$, where $\A$ is a linear degradation determined by the application, and $\w$ is white Gaussian noise.
Accordingly, the loss function is given by $f(\x) := \frac{1}{2} \lVert \A \x - \b \rVert_2^2$.

\begin{figure*}[t!]
\centering
\subfloat[Ground-truth.]{\includegraphics[width=0.126\linewidth]{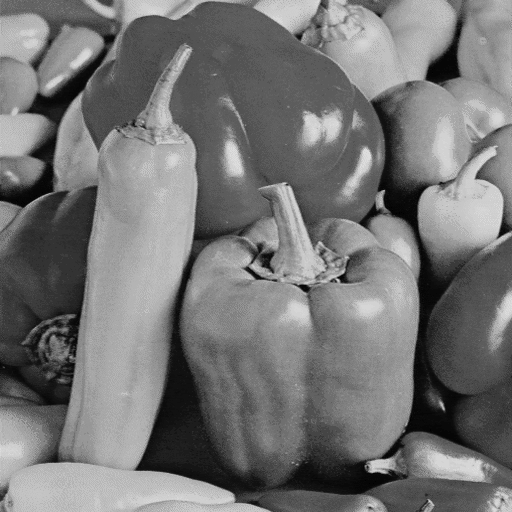}}
\hspace{0.1mm}
\subfloat[Observation.]{\includegraphics[width=0.126\linewidth]{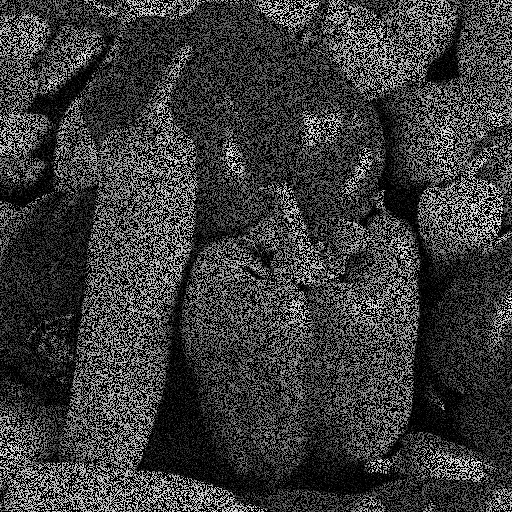}}
\hspace{0.1mm}
\subfloat[S-PnP-ADMM.]{\includegraphics[width=0.126\linewidth]{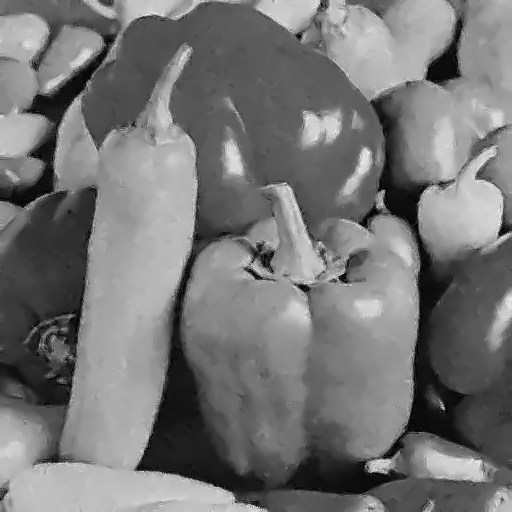}}
\hspace{0.1mm}
\subfloat[PnP-ADMM.]{\includegraphics[width=0.126\linewidth]{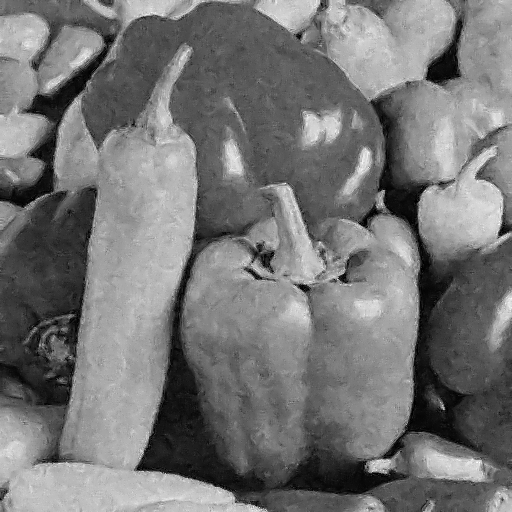}}
\hspace{0.1mm}
\subfloat[$f(\x_k) + g(\z_k)$.]{\includegraphics[height=0.126\linewidth]{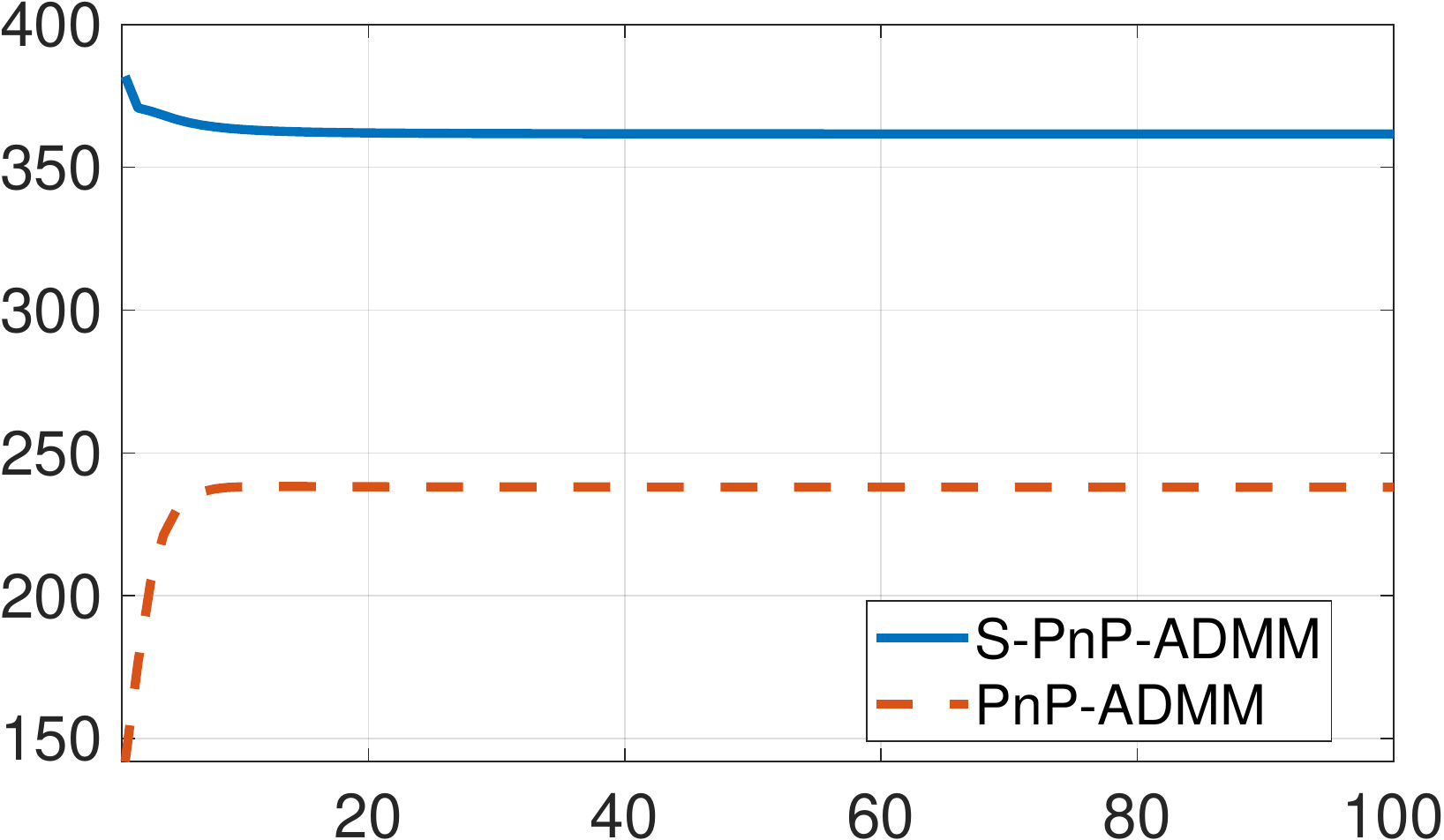}}
\hspace{0.1mm}
\subfloat[$\log \| \x_k - \z_k \|_2$.]{\includegraphics[height=0.126\linewidth]{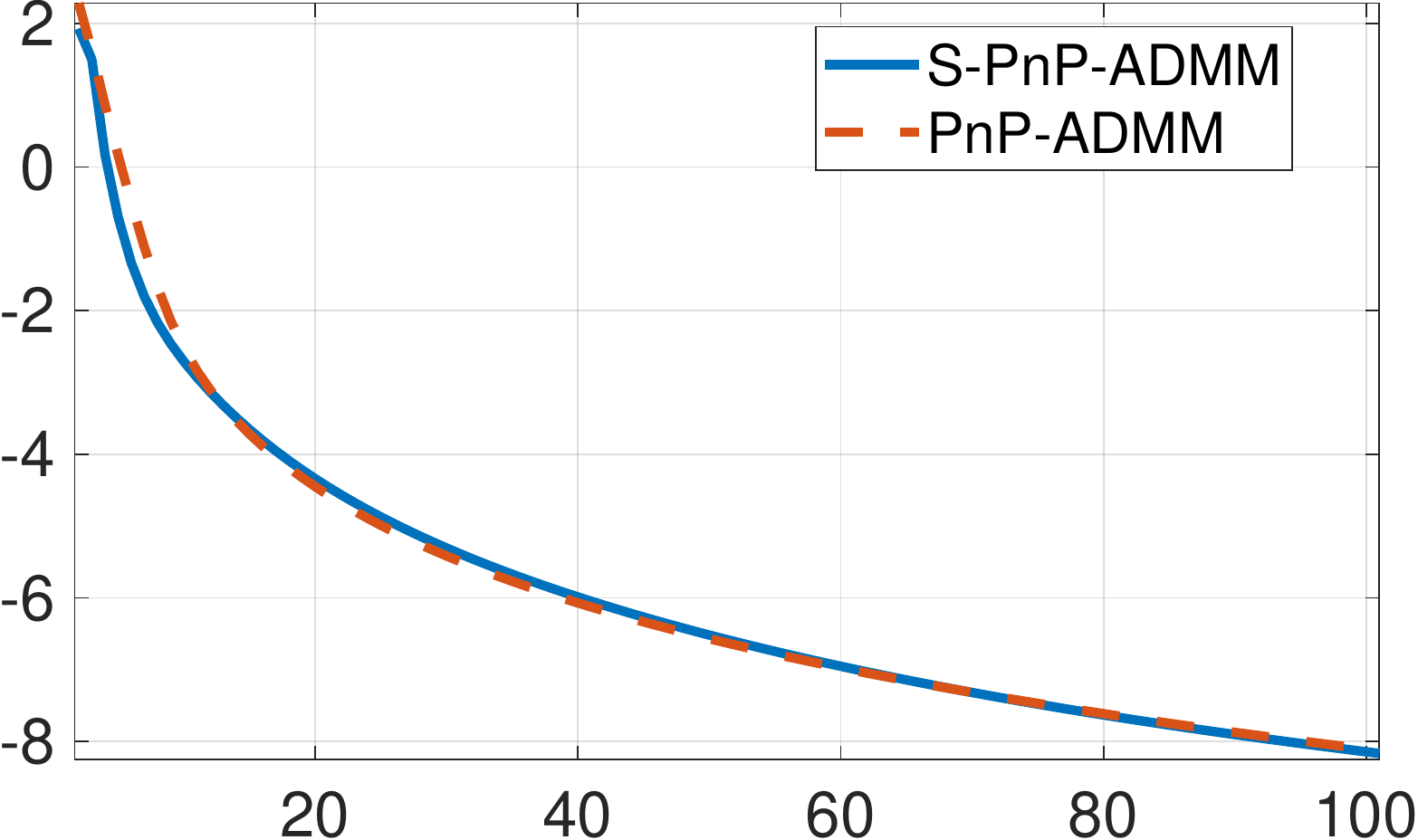}}
\caption{Example of image inpainting for the {\em peppers} image. (c) and (d): Restored images obtained using scaled PnP-ADMM (S-PnP-ADMM) and standard PnP-ADMM, respectively ($\rho=1$ in both cases). The respective PSNRs w.r.t. the ground-truth image are $29.04\,\si{dB}$ and $27.98\,\si{dB}$. (e): Plots of the objective values, $f(\x_k) + g(\z_k)$ vs $k$. (f): Plots of the residual values on a log scale, $\log \| \x_k - \z_k \|_2$ vs $k$.}
\label{fig:inpainting}
\end{figure*}

We use NLM as the denoiser in the scaled PnP algorithms.
Ideally, NLM denoises a pixel by taking the weighted average of all other pixels in the image, where the weights are determined by a Gaussian function applied to patch affinities \cite{Buades2005_NLM}.
However, in order to reduce computations, the practical implementation of NLM typically uses
only a neighborhood around each pixel for averaging.
Note that this no longer ensures the positive semidefiniteness of the kernel matrix $\K$; fortunately, this issue can be resolved as explained below (see \cite{Sreehari2016_PnP} for more details). 
If the neighborhood pixels are spatially weighted by a locally supported hat function (rather than a box function) centered on the current pixel, then the resulting $\K$ is positive semidefinite \cite[Appendix B]{Sreehari2016_PnP}.
In other words, one can use the following kernel function to construct $\K \in \Sy^n_+$:
\begin{equation*}
\kappa(\i,\j) = \eta(\i - \j) \varphi(P_{\i} - P_{\j}),
\end{equation*}
where $\eta$ is the two-dimensional hat function with width equal to the neighborhood size, $P_{\i}$ denotes a patch around pixel $\i$, and $\varphi$ is a multivariate Gaussian function with a fixed variance.
Consequently, Corollary \ref{cor:kernel} also applies to the above practical implementation of NLM.

To gain insight into the behavior of scaled PnP algorithms, we compare their convergence behavior with their standard PnP counterparts.
We use the symmetric DSG-NLM denoiser (which is derived from NLM) in the standard PnP algorithms.
Recall from \cite{Sreehari2016_PnP} that DSG-NLM is the proximal map of a convex function.
Thus, for standard PnP-FISTA and standard PnP-ADMM, both using DSG-NLM, the convergence properties in Theorems \ref{thm:FISTA_conv} and \ref{thm:ADMM_conv}, respectively, hold.

For our experiments, we use a neighborhood of size $11 \times 11$ and a patch size of $7 \times 7$ in both NLM and DSG-NLM.
Both the denoisers are implemented using fast filtering algorithms \cite{Darbon2008,Unni2018_PnP_LADMM}.
Though this is not the primary aim of the experiments, we additionally show that the restoration quality of scaled PnP algorithms is comparable to (and at times better than) their standard PnP counterparts.
Moreover, scaled PnP algorithms have faster runtimes.
We use a dataset of twenty images to evaluate the performance of the two algorithms.
The 20-image set consists of the standard Set12 dataset\footnote{Available at: \url{https://github.com/cszn/DnCNN/tree/master/testsets/Set12}.} (containing 12 images), along with eight commonly used test images\footnote{Downloaded from: \url{http://imageprocessingplace.com/root_files_V3/image_databases.htm}.} shown in Fig. \ref{fig:image_set}.
All the images are resized to $512 \times 512$.

\begin{figure*}[t!]
\centering
\subfloat[Ground-truth.]{\includegraphics[width=0.126\linewidth]{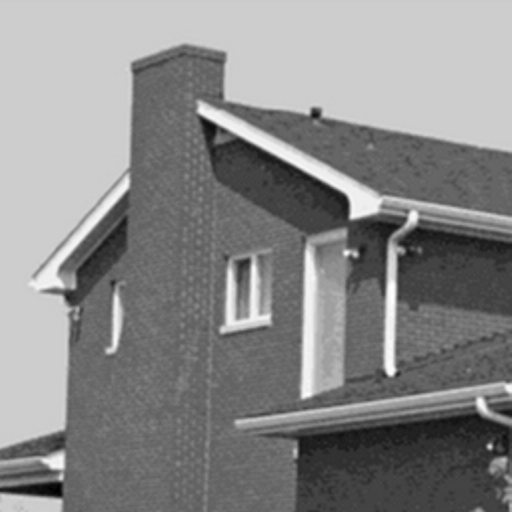}}
\hspace{0.1mm}
\subfloat[Observation.]{\includegraphics[width=0.126\linewidth]{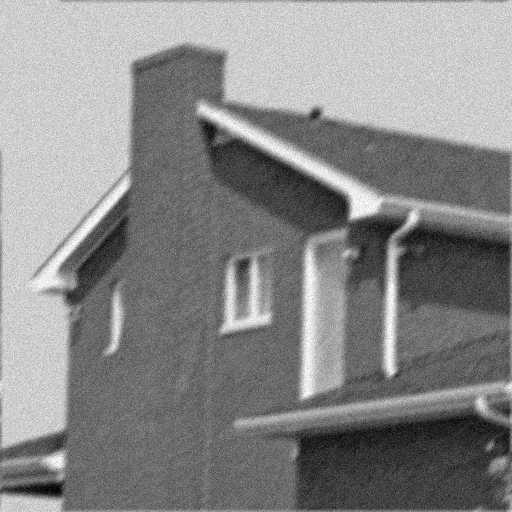}}
\hspace{0.1mm}
\subfloat[S-PnP-FISTA.]{\includegraphics[width=0.126\linewidth]{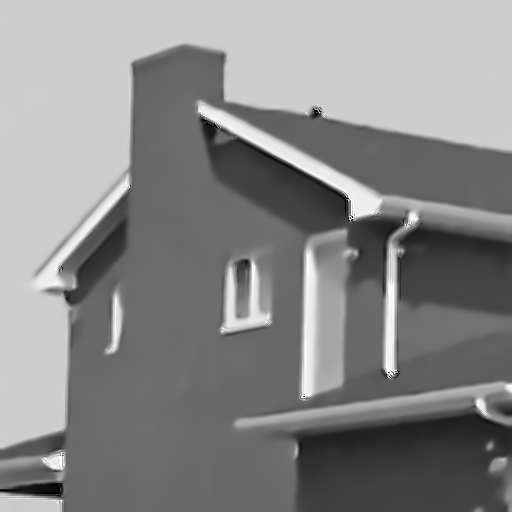}}
\hspace{0.1mm}
\subfloat[PnP-FISTA.]{\includegraphics[width=0.126\linewidth]{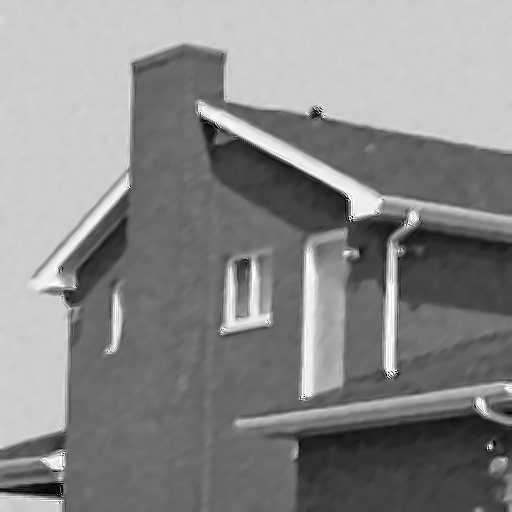}}
\hspace{0.1mm}
\subfloat[$f(\x_k) + g(\x_k)$.]{\includegraphics[height=0.126\linewidth]{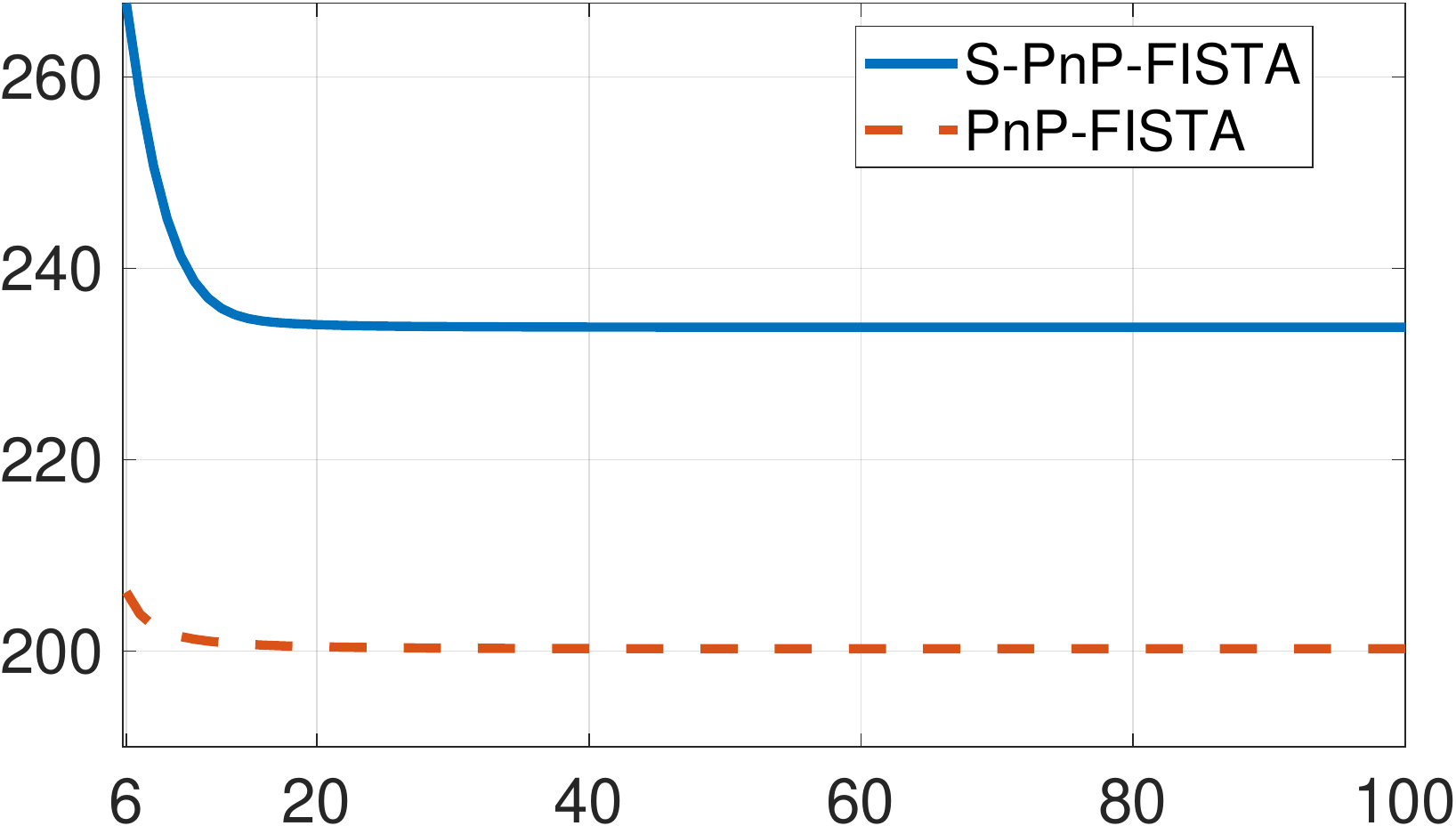}}
\hspace{0.1mm}
\subfloat[$\log \| \x_{k+1} - \x_k \|_2$.]{\includegraphics[height=0.126\linewidth]{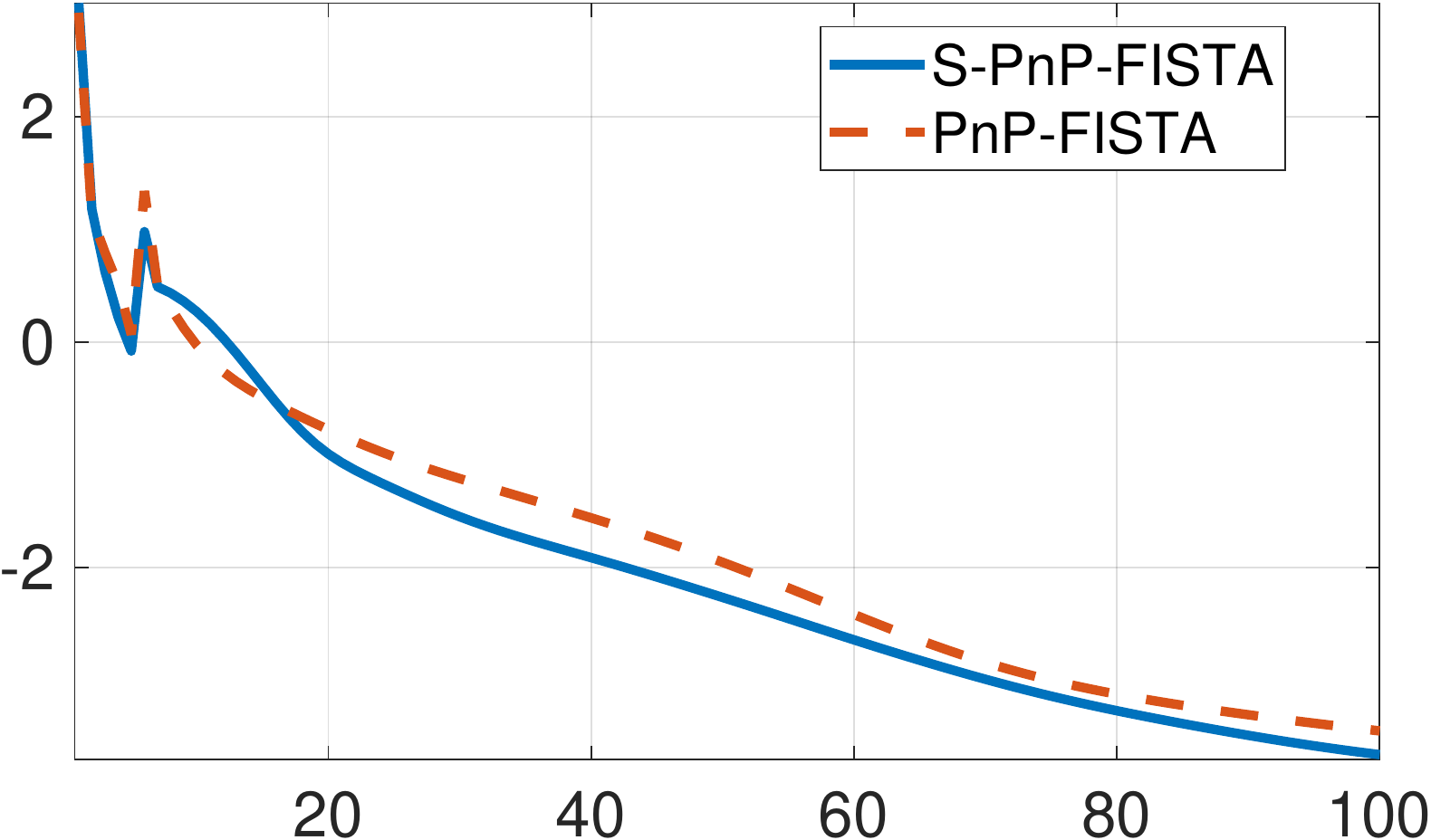}}
\caption{Example of motion deblurring for the {\em house} image. (c) and (d): Restored images obtained using scaled PnP-FISTA (S-PnP-FISTA) and standard PnP-FISTA, respectively. The respective PSNRs w.r.t. the ground-truth image are $31.28\,\si{dB}$ and $31.15\,\si{dB}$. (e): Plots of the objective values, $f(\x_k) + g(\x_k)$ vs $k$. (f): Plots of $\log \| \x_{k+1} - \x_k \|_2$ vs $k$.}
\label{fig:deblurring}
\end{figure*}

Note that in some of our results (see Fig.\,\ref{fig:inpainting}(e) and Fig.\,\ref{fig:deblurring}(e)) we evaluate the objective value $f + g$ in each iteration.
This requires evaluating $\Phi$ from \eqref{eq:phi}.
Since $\Phi$ is defined in terms of an eigendecomposition of $\W$, which is computationally difficult due to its huge size, it is not obvious how $\Phi$ can be efficiently evaluated.
We explain how this can be done in Appendix-\ref{subsec:obj_fun}.
The Matlab code of our implementation is available at \cite{sourcecode}.

\subsection{Image Inpainting}
In image inpainting, $\A \in \Re^{m \times n}$ is obtained by selecting $m < n$ rows of $\I \in \Re^{n \times n}$.
To estimate the ground-truth image, we use standard PnP-ADMM and scaled PnP-ADMM.
We set $\bnu_1 = \ZE$ and $\z_1$ to be the image obtained by applying a median filter to $\A^\top \b$ \cite{Tirer2019_iter_denoising}.
The denoiser $\W$ is calculated from $\z_1$ and kept fixed in subsequent iterations.
It follows from Definition \ref{def:scaled-prox} and \eqref{eq:ADMM_scaled_prox_x} that in scaled PnP-ADMM,
\begin{equation}
\label{inv}
\x_{k+1} = (\A^\top \A + \rho \H)^{-1} \big( \A^\top \b + \rho \H (\z_k - \rho^{-1}\bnu_k) \big).
\end{equation}
Since $\A^\top \A$ and $\H$ are diagonal, the inversion in \eqref{inv} reduces to elementwise division.

\begin{table}[t!]
\centering
\caption{Image inpainting: Average PSNRs (in \si{dB}) of scaled/standard PnP-ADMM across twenty images.}
\label{tab:inpainting}
\begin{tabular}{|c|ccc|}
\hline
\diagbox{$\sigma_{\w}$}{$m/n$} & $0.3$ & $0.5$ & $0.7$ \\
\hline
$10/255$ & $28.29/27.86$ & $30.36/30.03$ & $31.96/20.68$ \\
$20/255$ & $27.35/26.83$ & $28.88/27.32$ & $29.85/28.86$ \\
$30/255$ & $26.23/25.40$ & $27.51/26.16$ & $28.47/27.26$ \\
\hline
\end{tabular}
\end{table}

In Fig.\,\ref{fig:inpainting}, we show inpainting results for the {\em peppers} image, where $m/n = 0.5$ (i.e., $50\%$ missing pixels) and the standard deviation of noise is $\sigma_{\w} = 20/255$.
From Fig.\,\ref{fig:inpainting}(e) and Fig.\,\ref{fig:inpainting}(f), note that $f(\x_k) + g(\z_k)$ in scaled PnP-ADMM attains a stable value, whereas $\| \x_k - \z_k \|_2$ decays to $0$.
These observations agree with Theorem \ref{thm:ADMM_scaled_conv}.
We also plot the respective quantities for standard PnP-ADMM; observe that the objective value of standard PnP-ADMM increases before attaining a stable value.
Note that in ADMM, the iterates are not guaranteed to be in the feasible set; hence, it is possible for $f(\x_k) + g(\z_k)$ to take values less than the optimal value.
Also, note that the optimal values are different for the two algorithms since they use different regularizers.
In terms of PSNR (see the caption of Fig.\,\ref{fig:inpainting}) and visual quality, it is observed that the restored image is better for scaled PnP-ADMM.

We repeat the above experiment for $m/n \in \{0.3,0.5,0.7\}$ and $\sigma_{\w} \in \{10/255,20/255,30/255\}$ on the twenty images described earlier.
The average PSNRs across all images are reported in Table \ref{tab:inpainting} (the internal parameters were tuned to obtain the best PSNR).

\begin{figure}[t!]
\centering
\subfloat[NLM.]{\includegraphics[width=0.33\linewidth]{inpainting_mod_NLM.png}}
\hfill
\subfloat[TV.]{\includegraphics[width=0.33\linewidth]{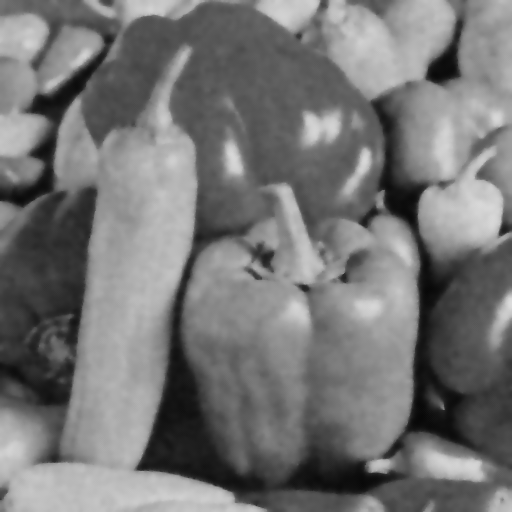}}
\hfill
\subfloat[DnCNN.]{\includegraphics[width=0.33\linewidth]{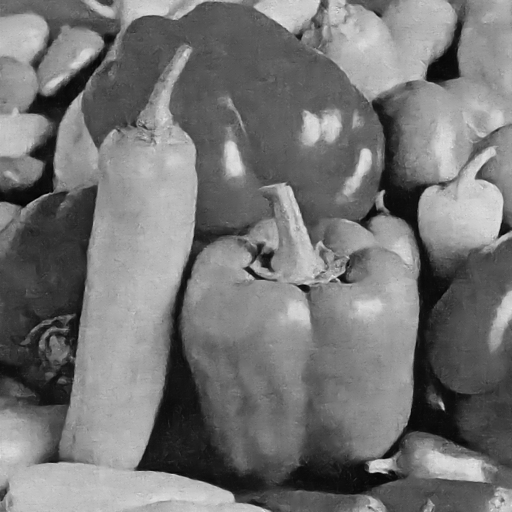}}
\caption{Restored images using three PnP methods: (a) Scaled PnP-ADMM using NLM denoiser (same as in Fig. \ref{fig:inpainting}(c)): (b) Standard PnP-ADMM with TV denoiser; (c) Standard PnP-ADMM with DnCNN denoiser. The PSNRs are (a) $29.04$ \si{dB}, (b) $27.87$ \si{dB}, and (c) $30.00$ \si{dB}. The input image is the same as in Fig. \ref{fig:inpainting}.}
\label{fig:inpainting_nonlinear}
\end{figure}

We have observed that in practice, the iterates of scaled PnP-ADMM typically stabilize in about $30$ to $50$ iterations; this observation is consistent with Theorem \ref{thm:ADMM_scaled_conv}.
Moreover, our observation is that scaled PnP-ADMM generally yields higher PSNRs than standard PnP-ADMM for this application.
Another advantage of using scaled PnP-ADMM is the reduced computational cost per iteration.
Indeed, it is known that the cost of DSG-NLM is about three times that of  NLM \cite{Unni2018_PnP_LADMM}.
This is reflected in the average per-iteration runtimes (in Matlab) of scaled PnP-ADMM and standard PnP-ADMM, which are $2.5\,\si{s}$ and $7.0\,\si{s}$, respectively, on our system ($2\,\si{GHz}$ processor with $8$ cores, $8\,\si{GB}$ memory).

For completeness, we also compare the restoration quality of the image in Fig. \ref{fig:inpainting}(c) with standard PnP-ADMM using a couple of nonlinear denoisers -- total variation (TV) denoising and DnCNN.
The restored images obtained using all three methods are shown in Fig. \ref{fig:inpainting_nonlinear}, where the input image is the same as that in Fig. \ref{fig:inpainting}.
Note that out of the three denoisers, TV yields the lowest PSNR, whereas DnCNN yields the highest PSNR.
This trend is also seen in the average PSNR values over the $20$ test images for the problem settings in question ($m/n=0.5$, $\sigma_{\w} = 20/255$) -- $28.88$ dB, $27.19$ dB and $29.18$ dB for NLM, TV and DnCNN, respectively.
Indeed, this is not surprising since it is well-known that in PnP algorithms, DnCNN (and other deep denoisers) empirically perform better than traditional denoisers such as NLM.

\subsection{Image Deblurring}
In image deblurring, $\A \in \Re^{n \times n}$ is a circulant blur matrix that convolves the input image with a known Point Spread Function (PSF).
We use standard PnP-FISTA and scaled PnP-FISTA to estimate the ground-truth image.
Note that $f$ is $\varepsilon$-smooth (see Definition \ref{def:rho-smooth}), where $\varepsilon$ is the largest eigenvalue of $\A^\top \A$; it can be computed using the discrete Fourier transform.
Furthermore, it is not difficult to show that $f$ is $\rho$-smooth with respect to $\H \in \Sy_{++}^n$ (see \eqref{eq:rho-H-smooth}), where $\rho = \varepsilon \lVert \H^{-1} \rVert_2$ and $\lVert \H^{-1} \rVert_2$ is the spectral norm of $\H^{-1}$.
This way one can determine a value for $\rho$ for standard PnP-FISTA and scaled PnP-FISTA.
We set $\x_0 = \b$ and freeze $\W$ after $5$ iterations (as discussed in Section \ref{subsec:kernel_den}).

In Fig.\,\ref{fig:deblurring}, we show results for the {\em house} image, where $\A$ is an operator that performs motion blurring using a $11 \times 11$ PSF, and $\sigma_{\w} = 10/255$.
The plot in Fig.\,\ref{fig:deblurring}(e) shows the evolution of $f(\x_k) + g(\x_k)$ with $k$.
As stated in Theorem \ref{thm:FISTA_scaled_conv}, the objective seems to converge to the optimal value.
Note that the objective function is well-defined once $\W$ is fixed, i.e., for $k \geq 6$.
For comparison, notice that the objective value of standard PnP-FISTA also stabilizes.
Moreover, as evident from Fig.\,\ref{fig:deblurring}(f), the successive difference $\| \x_{k+1} - \x_k \|_2$ decays to $0$ in both the versions (standard and scaled) of PnP-FISTA.
The PSNRs of the output images of the two algorithms are similar (see the caption of Fig.\,\ref{fig:deblurring}); however, the output of scaled PnP-FISTA is smoother than that of standard PnP-FISTA. In fact, we have observed that this is usually the case with most experiments.

We repeat the experiment for three choices of PSFs -- box PSF ($9 \times 9$), Gaussian PSF (variance $2^2$) and motion blurring PSF (as above); we vary the noise standard deviation $\sigma_{\w}$ over $\{5/255,10/255,15/255\}$.
In Table \ref{tab:deblurring}, we report average PSNRs across the twenty images in our dataset.
We observed that standard PnP-FISTA yields slightly better results in terms of PSNR; but for high noise level ($\sigma_{\w} = 15/255$), the PSNRs of the two PnP algorithms are similar. 
As before, due to the lower computational cost of NLM, scaled PnP-FISTA is faster ($2.0\,\si{s}$ per iteration) than standard PnP-FISTA ($5.5\,\si{s}$).

\begin{table}[t!]
\centering
\caption{Image deblurring: Average PSNRs (in \si{dB}) of scaled/standard PnP-FISTA across twenty images.}
\label{tab:deblurring}
\begin{tabular}{|c|ccc|}
\hline
\diagbox{$\sigma_{\w}$}{PSF} & Box & Gaussian & Motion \\
\hline
$5/255$ & $27.19/27.72$ & $28.98/29.62$ & $27.31/27.82$ \\
$10/255$ & $26.53/26.84$ & $27.93/28.60$ & $26.57/27.06$ \\
$15/255$ & $26.00/26.36$ & $27.33/27.38$ & $26.07/26.01$ \\
\hline
\end{tabular}
\end{table}

\section{Conclusion}
\label{sec:conc}
We provided a useful characterization for a linear denoiser to be a scaled proximal map, where the associated scaling matrix is naturally induced by the denoiser.
This subsumes existing results on symmetric positive semidefinite denoisers; importantly, it also applies to non-symmetric kernel denoisers. 
In keeping with this result, we introduced certain modifications in PnP-FISTA and PnP-ADMM.
We proved that the modified PnP algorithms (scaled PnP-FISTA and scaled PnP-ADMM), when used with the denoiser in question, essentially minimize an objective of the form $f + g$.
This resolves the question of optimality (and convergence) of PnP algorithms for a wider class of linear denoisers than was known till date.
To the best of our knowledge, this connection between scaled proximal maps and PnP regularization is novel.
We numerically validated our convergence results for the modified algorithms and showed that they produce compelling results for image restoration.
Note that in principle, the proposed modifications can also be applied, while ensuring convergence, to PnP iterations based on other proximal algorithms \cite{Parikh2014_prox} and splitting methods \cite{Eckstein1992_DouglasRachford}, \cite[Chaps. 26 and  28]{Bauschke2017_convex}.

\appendix
The following lemma is used in the proofs of Theorem \ref{thm:FISTA_scaled_conv} and Theorem \ref{thm:ADMM_scaled_conv}.
\begin{lemma}
\label{lem:prox}
Let $g \colon \Re^n \to \Rc$ be a closed proper convex function, $\H \in \mathbb{S}_{++}^n$ and $\gamma := g \circ \H^{-\frac{1}{2}}$.
Then 
\begin{equation*}
\label{eq:prox_relation}
\prox_{g,\normH{\cdot}}(\y) = \H^{-\frac{1}{2}} \ \prox_{\gamma,\normI{\cdot}}(\H^{\frac{1}{2}} \y) \qquad \forall \y \in \Re^n.
\end{equation*}
\end{lemma}
\begin{IEEEproof}
Take an arbitrary $\y \in \Re^n$; let $\x^\star = \prox_{g,\normH{\cdot}}(\y)$. It follows from Definition \ref{def:scaled-prox} that for all $\x \in \Re^n$,
\begin{equation}
\label{eq:prox gHy}
\frac{1}{2} \normH{\x^\star - \y}^2 + g(\x^\star) \leq \frac{1}{2} \normH{\x - \y}^2 + g(\x).
\end{equation}
Note that $g = \gamma \circ \H^{\frac{1}{2}}$; also, note that for $\u \in \Re^n$, we have $\normH{\u} = \| \H^{\frac{1}{2}} \u\|_2$. Therefore, using the change of variable $\z = \H^{\frac{1}{2}} \x$, we can rewrite \eqref{eq:prox gHy} as follows: for all $\z \in \Re^n$,
\begin{equation*}
\frac{1}{2} \|\H^{\frac{1}{2}} \x^\star - \H^{\frac{1}{2}} \y \|_2^2 + \gamma(\H^{\frac{1}{2}} \x^\star) \leq \frac{1}{2} \| \z - \H^{\frac{1}{2}} \y \|_2^2 + \gamma(\z).
\end{equation*}
As a result, $\H^{\frac{1}{2}} \x^\star = \prox_{\gamma,\normI{\cdot}}(\H^{\frac{1}{2}} \y)$.
Therefore,
\begin{equation*}
\prox_{g,\normH{\cdot}}(\y) = \H^{-\frac{1}{2}} \ \prox_{\gamma,\normI{\cdot}}(\H^{\frac{1}{2}} \y) \qquad \forall \y \in \Re^n. \IEEEQEDhereeqn
\end{equation*}
\end{IEEEproof}

\subsection{Proof of Theorem \ref{thm:FISTA_scaled_conv}}
\label{subsec:proof_FISTA_scaled_conv}
\begin{IEEEproof}
By Theorem \ref{thm:main}, we have $\mcD = \prox_{\Phi,\normH{\cdot}}$, where $\Phi \colon \Re^n \to \Rc$ is a closed proper convex function given by \eqref{eq:phi}. 
Furthermore, since $g= \rho \Phi$, the updates \eqref{eq:FISTA_scaled_t}\,-\,\eqref{eq:FISTA_scaled_x} in scaled PnP-FISTA can be written as follows: 
\begin{subequations}
\begin{align}
t_{k+1} &= \frac{1}{2} \left( 1 + \sqrt{1 + 4 t_k^2} \right), \label{eq:proof_FISTA_t}\\
\y_{k+1} &= \x_k + \frac{t_k - 1}{t_{k+1}} (\x_k - \x_{k-1}), \label{eq:proof_FISTA_y}\\
\x_{k+1} &= \prox_{g/\rho,\normH{\cdot}} \big(\y_{k+1} - \rho^{-1}\H^{-1} \nabla\!f(\y_{k+1})\big). \label{eq:proof_FISTA_x}
\end{align}
\end{subequations}
Let $\beta := f \circ \H^{-\frac{1}{2}}$ and $\gamma := g \circ \H^{-\frac{1}{2}}$. 
Note that as $\rho >0$, it follows from \eqref{eq:phi} and the definitions of functions $g$ and $\gamma$ that $\gamma \colon \Re^n \to \Rc$ is of the form $\gamma = \alpha + i_C $, where $\alpha \colon \Re^n \to \Re$ is a convex quadratic function, and $i_C$ is the indicator function of $C = \big\{\H^{\frac{1}{2}} \x : \x \in \mcR(\W)\big\}$. 

Next, we show that $\beta: \Re^n \to \Re$ is convex and $\rho$-smooth.
As $f$ is a differentiable convex function, it follows from the definition of $\beta$ that it is a differentiable convex function; moreover, by the chain rule,
\begin{equation}
\label{eq:grad-beta}
\nabla\!\beta(\v) = \H^{-\frac{1}{2}} \nabla\!f(\H^{-\frac{1}{2}} \v) \qquad \forall \v \in \Re^n.
\end{equation}
Recall that for $\u \in \Re^n$, we have $\normI{\u} = \|\H^{-\frac{1}{2}} \u\|_{\H}$.
Now, since $f$ is $\rho$-smooth with respect to $\H \in \Sy^n_{++}$, it follows from \eqref{eq:rho-H-smooth} and \eqref{eq:grad-beta} that for all $\v_1, \v_2 \in \Re^n$,
\begin{equation*}
\|\nabla\!\beta(\v_1) - \nabla\!\beta(\v_2)\|_2 \,\leq\, \rho\, \|\v_1 - \v_2 \|_2. 
\end{equation*}
Now, consider the following convex program:
\begin{equation}
\label{eq:proof_FISTA_prob}
\minimize_{\u \in \Re^n}\; \beta(\u) + \gamma(\u).
\end{equation}
It follows from the definition of functions $\beta$ and $\gamma$ that the optimal values of \eqref{eq:proof_FISTA_prob} and \eqref{eq:main_prob} are equal.
Furthermore, $\x^\star$ is a minimizer of \eqref{eq:main_prob} if and only if $\u^\star = \H^{\frac{1}{2}} \x^\star$ is a minimizer of \eqref{eq:proof_FISTA_prob}. 
Note that problem \eqref{eq:proof_FISTA_prob} satisfies the conditions given in Theorem \ref{thm:FISTA_conv}.

Using Lemma \ref{lem:prox} and \eqref{eq:grad-beta}, we can rewrite \eqref{eq:proof_FISTA_t}\,-\,\eqref{eq:proof_FISTA_x} as 
\begin{subequations}
\begin{align}
t_{k+1} &= \frac{1}{2} \left( 1 + \sqrt{1 + 4 t_k^2} \right), \label{eq:proof_FISTA_t-new}\\
\v_{k+1} &= \u_k + \frac{t_k - 1}{t_{k+1}} (\u_k - \u_{k-1}), \label{eq:proof_FISTA_v}\\
\u_{k+1} &= \prox_{\gamma/\rho,\normI{\cdot}}\big(\v_{k+1} - \rho^{-1}\nabla\!\beta(\v_{k+1})\big), \label{eq:proof_FISTA_u}
\end{align}
\end{subequations}
where $\u_k = \H^{\frac{1}{2}} \x_k$ and $\v_k = \H^{\frac{1}{2}} \y_k$ for $k \in \Na$.
By applying Theorem \ref{thm:FISTA_conv} to \eqref{eq:proof_FISTA_prob}, we can conclude that the iterates \eqref{eq:proof_FISTA_t-new}\,-\,\eqref{eq:proof_FISTA_u} satisfy the following:
\begin{equation*}
\beta(\u_k) + \gamma(\u_k) \leq p^\star + O(1/k^2),
\end{equation*}
whereby \eqref{eq:conv1} follows immediately.
\end{IEEEproof}

\subsection{Proof of Theorem \ref{thm:ADMM_scaled_conv}}
\label{subsec:proof_ADMM_scaled_conv}
\begin{IEEEproof}
By Theorem \ref{thm:main}, we have $\mcD = \prox_{\Phi,\normH{\cdot}}$, where $\Phi \colon \Re^n \to \Rc$ is a closed proper convex function given by \eqref{eq:phi}. 
Furthermore, since $g = \rho \Phi$, the updates \eqref{eq:ADMM_scaled_prox_x}\,-\,\eqref{eq:ADMM_scaled_prox_nu} in scaled PnP-ADMM can be written as follows: 
\begin{subequations}
\begin{align}
\x_{k+1} &= \prox_{f/\rho,\normH{\cdot}}(\z_k - \rho^{-1}\bnu_k), \label{eq:proof_ADMM_x}\\
\z_{k+1} &= \prox_{g/\rho,\normH{\cdot}}(\x_{k+1} + \rho^{-1}\bnu_k), \label{eq:proof_ADMM_z}\\
\bnu_{k+1} &= \bnu_k + \rho(\x_{k+1} - \z_{k+1}). \label{eq:proof_ADMM_nu}
\end{align}
\end{subequations}

Let $\beta := f \circ \H^{-\frac{1}{2}}$ and $\gamma := g \circ \H^{-\frac{1}{2}}$; note that $\beta$ and $\gamma$ are closed proper convex functions. Now, consider the following convex program:
\begin{mini}
{}{\beta(\tilde{\x}) + \gamma(\tilde{\z})}{\label{eq:proof_ADMM_prob}}{}
\addConstraint{\tilde{\x}}{=\tilde{\z.}}{}
\end{mini}
Let $\mcM$ be the Lagrangian of \eqref{eq:proof_ADMM_prob}, i.e.,
\begin{equation*}
\label{eq:Lagrangian-proof_ADMM_prob}
\mcM(\tilde{\x},\tilde{\z},\tilde{\bnu}) := \beta(\tilde{\x}) + \gamma(\tilde{\z}) + \tilde{\bnu}^\top (\tilde{\x}-\tilde{\z}).
\end{equation*}
It follows from the definitions of  $\beta$ and $\gamma$ that the optimal values of \eqref{eq:proof_ADMM_prob} and \eqref{eq:main_prob_const} are equal.
Furthermore, $(\x^\star,\z^\star,\bnu^\star)$ is a saddle point of $\mcL_{\H}$ given by \eqref{eq:scaled_Lagrangian-main_prob_const} if and only if $(\H^{\frac{1}{2}} \x^\star,\H^{\frac{1}{2}} \z^\star,\H^{\frac{1}{2}} \bnu^\star)$ is a saddle point of $\mcM$.
 
Using Lemma \ref{lem:prox}, we can rewrite \eqref{eq:proof_ADMM_x}\,-\,\eqref{eq:proof_ADMM_nu} as follows:
\begin{subequations}
\begin{align}
\tilde{\x}_{k+1} &= \prox_{\beta/\rho,\normI{\cdot}}(\tilde{\z}_k - \rho^{-1}\tilde{\bnu}_k), \label{eq:proof_ADMM_x-tilde}\\
\tilde{\z}_{k+1} &= \prox_{\gamma/\rho,\normI{\cdot}}(\tilde{\x}_{k+1} + \rho^{-1}\tilde{\bnu}_k), \label{eq:proof_ADMM_z-tilde}\\
\tilde{\bnu}_{k+1} &= \tilde{\bnu}_k + \rho(\tilde{\x}_{k+1} - \tilde{\z}_{k+1}) \label{eq:proof_ADMM_nu-tilde},
\end{align}
\end{subequations}
where $\tilde{\x}_k = \H^{\frac{1}{2}} \x_k$, $\tilde{\z}_k = \H^{\frac{1}{2}} \z_k$ and $\tilde{\bnu}_k = \H^{\frac{1}{2}} \bnu_k$ for $k \in \Na$.
On applying Theorem \ref{thm:ADMM_conv} to \eqref{eq:proof_ADMM_prob}, we can conclude that for arbitrary $\tilde{\z}_1, \tilde{\bnu}_1 \in \Re^n $, the iterates \eqref{eq:proof_ADMM_x-tilde}\,-\,\eqref{eq:proof_ADMM_nu-tilde} converge to a saddle point of $\mcM$, and $\lim_{k \to \infty} \big( \beta(\tilde{\x}_k) + \gamma(\tilde{\z}_k) \big) = p^\star$. 
Consequently, the iterates \eqref{eq:proof_ADMM_x}\,-\,\eqref{eq:proof_ADMM_nu} converge to a saddle point of $\mcL_{\H}$, and $\lim_{k \to \infty} \big(f(\x_k) + g(\z_k)\big) = p^\star$.
\end{IEEEproof}

\subsection{Divergence of Standard PnP-ADMM}
\label{subsec:counterexample}
The following example shows that standard PnP-ADMM with a kernel denoiser can fail to converge  (in the sense of Theorem \ref{thm:ADMM_conv}); note that the convergence of iterates in Theorem \ref{thm:ADMM_conv} imply that $\lim_{k \to \infty}\, (\x_k - \z_k) = \ZE$.
This example substantiates the necessity of the algorithmic modifications introduced in Section \ref{sec:pnp}.
\begin{example}
\label{ex:counterexample}
Let $\a^\top = \begin{bmatrix} 0.8295 & -0.5586 \end{bmatrix}$, $b = 1$, and $f \colon \Re^2 \to \Re$ be given by $f(\x) := \frac{1}{2}\,|\a^\top\x - b|^2$.
Note that $f$ is a closed proper convex function.
Let $\W = \D^{-1} \K$, where
\begin{equation*}
\D = \begin{bmatrix}
0.3116 & 0\\
0 & 0.5788
\end{bmatrix} \;\mbox{and}\; \K = \begin{bmatrix}
0.1102 & 0.2014\\
0.2014 & 0.3774
\end{bmatrix} \in \Sy^2_+.
\end{equation*}
By construction, $\W$ is of the form given in Corollary \ref{cor:kernel}.
Therefore, it follows from the proof of Corollary \ref{cor:kernel} that $\W$ is similar to $\D^{-\frac{1}{2}} \K \D^{-\frac{1}{2}} \in \Sy^2_+$ and $\lambdamax(\W) \leq 1$; in fact, $\lambdamax(\W) = 1$.
Furthermore, $\W$ is the $\D$-scaled proximal map of a closed proper convex function $\Phi$ of the form \eqref{eq:phi}.

Consider the standard PnP-ADMM algorithm with $\mcD = \W$ and $\rho=1$.
Let $\C = \I+\a\a^\top$, and
\begin{equation*}
\R = \begin{bmatrix} \W \\ \I - \W \end{bmatrix},\; \S^\top = \begin{bmatrix} \C^{-1} & \I - \C^{-1}\end{bmatrix},\;  \d = b\, \R \C^{-1} \a.
\end{equation*}
Let $\u_k^\top = \begin{bmatrix} \z_k^\top & \bnu_k^\top \end{bmatrix}$.
For the PnP-ADMM iterates, obtained from \eqref{eq:ADMM_prox_x}\,-\,\eqref{eq:ADMM_prox_nu}, we can write
\begin{subequations}
\label{eq:residue_divergence}
\begin{equation}
\label{eq:counter_state-update}
\u_{k+1} = \R \S^\top \u_k + \d,
\end{equation}
\begin{equation}
\label{eq:counter_residue}
\x_{k+1} - \z_{k+1} = \T \u_k + \q,
\end{equation}
\end{subequations}
where $\T = \begin{bmatrix}(\I - \W) \C^{-1} & - \big(\W + \C^{-1} - \W \C^{-1}\big)
\end{bmatrix}$ and $\q = b\,(\I - \W) \C^{-1} \a$.
Let $\u_1 = \ZE$; thus, it follows from \eqref{eq:counter_state-update} that for $k=2,3,\ldots$, we have
\begin{equation}
\label{eq:counter_state-var}
\u_k = \sum_{j=0}^{k-2} (\R \S^\top)^j \d.
\end{equation}
The following assertions are not difficult to verify:
\begin{enumerate}[(i)]
\item $\R \S^\top \in \Re^{4 \times 4}$ is diagonalizable over $\Re$, and its dominant eigenvalue is greater than $1$.
\item $\d$ has a nonzero component along an eigenvector $\v_1$ corresponding to the dominant eigenvalue of $\R \S^\top$.
\item $\v_1 \notin \mcN(\T)$.
\end{enumerate}
As a result, it follows from an analysis similar to the power method (see \cite[Sec. 5.3]{Watkins2010_matrix_comp}) that
\begin{equation*}
\lim_{j \to \infty}\, \left\|\T (\R \S^\top)^j \d \right\|_2 = \infty.
\end{equation*}
Consequently, it follows from \eqref{eq:counter_residue} and \eqref{eq:counter_state-var} that the residual sequence $(\x_k-\z_k)_{k \in \Na}$ diverges; see Table \ref{tab:counterexample} for a numerical validation of this conclusion.
\begin{table}[H]
\centering
\caption{Values of $\log \| \x_k - \z_k \|_2$ for Example \ref{ex:counterexample}.}
\label{tab:counterexample}
\begin{tabular}{|c|c|c|c|c|c|}
\hline
$k=1$ & $k=200$ & $k=400$ & $k=600$ & $k=800$ & $k=1000$ \\
\hline
$-0.6743$ & $-0.1045$ & $3.7808$ & $7.6662$ & $11.5515$ & $15.4369$\\
\hline
\end{tabular}
\end{table}
\end{example}

\subsection{Computation of $\Phi$}
\label{subsec:obj_fun}
To compute the objective value in the $k$-th iteration of the scaled PnP algorithms, we need to evaluate $\Phi(\u_k)$, where $\u_k = \x_k$ in scaled PnP-FISTA and $\u_k = \z_k$ in scaled PnP-ADMM.
We see from \eqref{eq:FISTA_scaled_x} and \eqref{eq:ADMM_scaled_prox_z} that for both the algorithms, $\u_k$ is given by $\u_k = \W \q_k$ for some (known) $\q_k$.
In particular, since $\u_k \in \mcR(\W)$, it follows from \eqref{eq:phi} that
\begin{equation}
\label{eq:quad-form-1}
\Phi(\u_k) = \frac{1}{2}\ \u_k^\top \P \u_k = \frac{1}{2}\ \q_k^\top \big(\W^\top \P \W\big) \q_k.
\end{equation}
Now, recall the matrices $\V$ and $\LA_r$ defined in the proof of Theorem \ref{thm:main}.
By writing $\W$, $\P$ and $\H$ in terms of $\V$ and $\LA_r$, it is not difficult to verify that
\begin{equation}
\label{eq:quad-form-2}
\W^\top \P \W = (\I - \W)^\top \H \W.
\end{equation}
It follows from \eqref{eq:quad-form-1} and \eqref{eq:quad-form-2} that
\begin{equation}
\label{eq:final-form}
\Phi(\u_k) =  \frac{1}{2}\ (\q_k - \u_k)^\top \H \u_k.
\end{equation}
Note that $\u_k$ and $\q_k$ are known, and $\H$ is diagonal for the NLM denoiser (see Corollary \ref{cor:kernel}). Thus, $\Phi(\u_k)$ can be easily evaluated from \eqref{eq:final-form}.
A similar approach is used for standard PnP algorithms using the DSG-NLM denoiser.

\section*{Acknowledgment}\label{sec:ack}
We thank the Associate Editor and reviewers for their valuable comments and suggestions.

\bibliographystyle{IEEEtran}
\bibliography{citations}

\end{document}